\documentclass[a4paper]{article}
\usepackage{amssymb}
\usepackage{hyperref, url, paralist, graphicx, pdflscape, amsfonts, siunitx, gensymb}
\usepackage[sumlimits]{amsmath}
\usepackage{bm}
\DeclareMathOperator*{\argmin}{arg\,min}
\usepackage{cite}
\usepackage{nameref,hyperref}
\usepackage{rotating}
\begin{document}
\begin{flushleft}
{\Large
\textbf\newline{Assessing Site Effects and Geographic Transferability when Interpolating Point Referenced Spatial Data: A Digital Soil Mapping Case Study} 
}
\newline
\\
Benjamin R. Fitzpatrick\textsuperscript{1,2,3,*},
David W. Lamb\textsuperscript{4,2},
Kerrie Mengersen\textsuperscript{5,1,2,3},
\newline
\newline
\small 1 Mathematical Sciences School, Queensland University of Technology (QUT), Brisbane, QLD 4001, Australia
\\
\small 2 Cooperative Research Centre for Spatial Information (CRCSI), Carlton, VIC 3053, Australia 
\\
\small 3 Institute for Future Environments, Queensland University of Technology (QUT), Brisbane, QLD 4001, Australia
\\
\small 4 Precision Agriculture Research Group, University of New England, Armidale, NSW 2351, Australia
\\
\small 5 ARC Centre of Excellence for Mathematical and Statistical Frontiers, Queensland University of Technology (QUT), Brisbane, QLD 4001, Australia
\\
* E-mail: Corresponding b1.fitzpatrick@qut.edu.au
\end{flushleft}

\begin{abstract}
When making inferences concerning the environment, ground truthed data will frequently be available as point referenced (geostatistical) observations that are clustered into multiple sites rather than uniformly spaced across the area of interest.
In such situations, the similarity of the dominant processes influencing the observed data across sites and the accuracy with which models fitted to data from one site can predict data from another site provide valuable information for scientists seeking to make inferences from these data.
Such information may motivate a more informed second round of modelling of the data and also provides insight into the generality of the models developed and an indication of how these models may perform at predicting observations from other sites.
We have investigated the geographic transferability of site specific models and compared the results of using different implementations of site specific effects in models for data combined from two sites.
Since we have access to data on a broad collection of environmental characteristics that each held potential to aid the interpolation of our geostatistical response observations we have investigated these issues within the framework of a computationally efficient method for variable selection when the number of explanatory variables exceeds the number of observations.
We have applied Least Absolute Shrinkage Selection Operator (LASSO) regularized Multiple Linear Regression (MLR) as fitted by the computationally efficient Least Angle Regression algorithm.
The response variable in our case study, soil carbon, is of interest as a potential location for the sequestration of atmospheric carbon dioxide and for its positive contribution to soil health and fertility. 
Investigating both the necessity of incorporating site effects when modelling the data from multiple sites and the geographic transferability of site specific models as we have done here may provide valuable insight to scientists from a variety of fields seeking to interpolate geostatistical observations from multiple sites.
\end{abstract}

\section*{Introduction}
\label{s:intro}
Maps of soil carbon levels are useful for soil management \cite{AldanaJague2015} and carbon accounting \cite{McBratney2014}.
Predicting quantitative maps of soil characteristics from empirical data has been referred to as digital soil mapping \cite{Minasny2013, Minasny2014}.
Digital soil mapping approaches utilizing environmental variables as covariates have frequently been applied in soil carbon mapping \cite{Mueller2003, Barnes2003, Simbahan2006b, Miklos2010, ViscarraRossel2014, Xiong2014}.
Digital soil mapping has been characterized by limited numbers of geostatistical response observations \cite{Minasny2014} and much finer resolution geostatistical data and or full cover areal data on diverse collections of environmental characteristics of potential relevance as covariates for modelling the response.
As such, the methodological challenges of digital soil mapping bear marked similarities to those encountered in other fields where a set of `ground truthed' geostatistical observations or non-contiguous areal observations (plots or quadrats) is sought to be interpolated or extrapolated with the aid of other environmental data available across the area of interest.
Examples of this analysis task outside soil science include modelling above ground biomass in forests \cite{Lu2016} and semi-arid regions \cite{Eisfelder2012} along with species distribution modelling and biogeography \cite{Franklin2009}.
As the spatial extent of the desired modelling increases, rather than a uniform density of response observations across the entire space, often the data available are from multiple, smaller scale studies of a collection of sub-regions or sites.
This raises the important question of geographic transferability of the models developed and the subsequent methodological question of how to develop a model that both appropriately describes the available data and adequately facilitates geographic transferability.
\newline
\newline
In this paper we examine a canonical example of this situation with data from two spatially proximate but non-contiguous sites.
Our aim is to build models that use high resolution environmental data to interpolate the response observations and produce full cover predicted maps of the response at both sites.
With such data we also have the opportunity to both explore issues associated with the geographic transferability of site specific models and to compare different methods for incorporating site effects when modelling data combined from both sites.
Each approach is employed within a cross validation scheme with final predictions created by model averaging the models selected by that approach from the training sets.
\newline
\newline
Geographic (spatial) transferability is evaluated in terms of the accuracy with which models fitted to data from one site can predict the response observations at the other site from the observations of the covariates at this other site.
Comparing the identities and distributions of the covariates selected for these models provides useful context for interpreting these assessments.
We also compare of three approaches to modelling data combined from the two sites.
Firstly, we ignore the grouping of the data into sites by fitting models where covariates are assigned a single parameter across all observations within a training set and model averaging the results.
This approach uses covariates as constant effects across both sites.
Hereafter we refer to such effects as `global' effects and models that only utilize global effects of covariates as global effects models.
Secondly, we expand upon the above approach by fitting site specific models to the residuals from the model averaged predictions from the global effects models.  
We then correct the model averaged predictions from the global effects models with the model averaged predictions from the site specific models for the residuals from the global effects based predictions.
Thirdly, we consider both global and site specific effects of covariates together in a single model fitting procedure by supplying design matrices to the variable selection algorithm that include all covariates as both global effects and site specific effects for each site.
We compare these three approaches to modelling data from two sites in terms of the identities of the covariates selected and the accuracy with which the observed data are predicted.
\newline
\newline
We conduct all modelling within the Multiple Linear Regression (MLR) framework as modified by Least Absolute Shrinkage and Selection Operator (LASSO) \cite{Tibshirani1996,Tibshirani2011} regularization for four reasons.
Firstly, the MLR framework is conducive to the creation of site specific terms for covariates.
Secondly, LASSO regularized MLR mitigates the concerns arising from conducting MLR with sets of covariates among which collinearity exists.
Thirdly, the computational efficiency of the Least Angle Regression (LAR) \cite{Efron2004} algorithm for calculating LASSO solutions enables broad sets of potential covariates to be considered as candidates for inclusion in selected models.
As such, the LAR algorithm enables the consideration of global and site specific versions of linear main effects, pairwise interactions of linear main effects and polynomial terms for all covariates.
Fourthly, LASSO solutions shrink a subset of the regression coefficients to zero exactly thereby conducting variable selection.
This provides estimates of the importance of both the covariates associated with these regressions coefficients and the nature in which they are used, be it as global effects, site specific effects and or interaction or polynomial terms.
\newline
\newline
This paper is structured as follows.
We commence the methods section with descriptions of the study site and data collection then go on to describe the response and covariates.
In the statistical methods section we first describe the interpolation of the covariates to spatial neighbourhoods around the response observations then go on to outline our approach to construction of the design matrices to produce the various treatments of site information used in the models fitted.
The recentring and rescaling of the covariates necessary prior to utilising the LAR algorithm is then described.
In the following section we introduce LASSO regularized MLR and describe the cross validation scheme we have used to estimate the tuning parameter in the LASSO method.
Subsequently, we describe how we model averaged the predictions from the cross validation scheme.
In the following section we describe how we compare the different models fitted in terms of predictive accuracy and the covariates selected.
We conclude the methods section with descriptions of how we predicted full cover maps of \%SOC at the two study sites, a description of the software used in the analysis and the location of the repository containing the code written to complete this analysis.

\section*{Methods}
\label{s:meth}
\subsection*{Study Area}
\begin{figure}[h!]
\label{Fig:HS_Terrain}
\includegraphics[width = \textwidth]{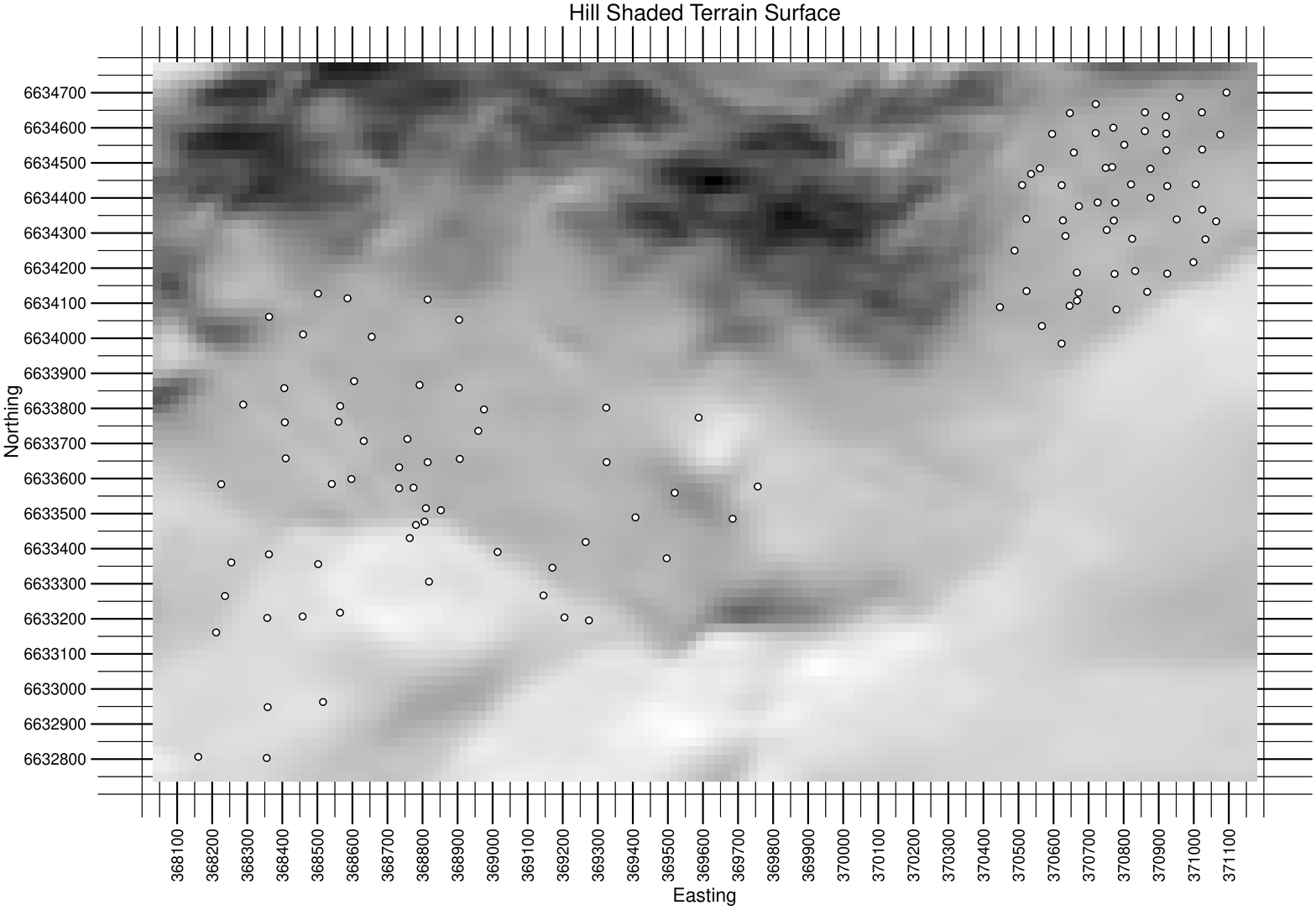}
\caption{Hillshaded terrain surface produced from the digital elevation model of Mt Duval and surrounds with soil core sample locations marked as white filled black circles. The more westerly group of soil cores form site B1 and the more easterly group of soil cores form site B2.}
\end{figure}

Our two study sites were situated on the Sustainable, Manageable, Accessible, Rural Technology (SMART) Farm of the University of New England near Armidale, New South Wales (NSW), Australia.
The SMART farm is located between coordinates \ang{30;22;59}S \ang{151;35;23}E (north-west corner) and \ang{30;27;26}S \ang{151;39;52}E (south-east corner).
Both sites were situated at the base of Mount Duval (1393m \cite{NationalParksandWildlifeService2003}) and consisted of a mixture of selectively cleared native pasture, remnant vegetation and regrowth.
The native pasture at both sites has a history of grazing by sheep and cattle.
Hereafter we will refer to the more westerly site as site B1 and the more easterly site as site B2.
These two sites are visible as the two groups of soil core sample locations superimposed as filled circles on the hillshaded terrain surface in Fig. \ref{Fig:HS_Terrain}.
Aerial photographs of the sites are presented as S1 Fig 1 and S2 Fig 2.
The soils of both study sites are constituents of the Uralla Plutonic Suite/Mount Duval Adamellite (acid porphyritic, hornblende-biotite monzogranite).
The hilltop soils of these sites typically consisted of yellow and brown chromosols \cite{Isbell2002} while the drainage routes were typically occupied by alluvial soils and siliceous sand complexes.
The SMART farm has summer dominated rainfall of approximately 790mm per year \cite{Garraway2011}. 

\subsection*{Data Collection}
The observations of our response variable were collected in-situ as soil cores which were then laboratory analysed for soil organic carbon content (\%SOC).
The other in-situ and remotely sensed data we were able to obtain for the study sites yielded 65 potential covariates.
The full names and acronyms for these covariates have been included in Table \ref{t:covar.acrons}.

\subsubsection*{In-situ Data}
An all terrain vehicle (ATV) drove transects across the two sites collecting top of pasture reflectance under active illumination and soil apparent electrical conductivity measurements in February, May and November of 2009.
The ATV towed a trailer equipped with a Geonics EM38 unit (Geonics Ontario Canada) which was operated in the vertical dipole orientation to measured the soil apparent electrical conductivity ($EC_a$).
The active illumination was provided by a Light Emitting Diode (LED) array and near-infrared (NIR) and visible red (RED) reflectance were measured by Crop Circle$^{TM}$ sensors from Holland Scientific, USA both of which were also mounted on the trailer.
The first such ATV survey was in February following a week of no rain in what was otherwise the second wettest month of the year, the second survey was conducted in May after a week of heavy rain and the third survey was conducted in November which marked the end of the winter growing season.
In addition to NIR and RED reflectance a selection of nine vegetation indices were calculated for each month (see Table\ref{t:covar.acrons}) as potential indicators of pasture biomass which in turn may have been correlated with soil organic carbon.
The differences between the ECA values from each of the three distinct pairing of months were also computed due to the potential these differences had to indicate changes in soil moisture between these months.
Changes in soil moisture may have been related to \%SOC levels via the influence SOC may exert on the infiltration of soil by water and the retention of water by soil\cite{Franzluebbers2002}.  
Soil organic carbon was measured by analysis of soil core samples with a Carlo Erba NA 1500 solid sample analyzer (Carlo Erba Instruments, Milan, Italy) \cite{Garraway2011}.
Soil cores were collected to a depth of 200mm.
Further information regarding the collection, preparation and analysis of soil cores may be found in Garraway et al. 2011.
Locations for soil core sampling were chosen by random sampling within each of five landscape functional types defined by k-means clustering the red, green and blue channels from the aerial imagery and the $EC_a$ data from the February ATV survey \cite{Garraway2011}.
A minimum of six locations at which to collect soil core samples within each landscape functional type were randomly selected with additional locations selected by Garraway et al. to better represent the landscape attributes of the study area.
The locations at which the soil cores were collected were georeferenced using a differential Global Positioning System (dGPS) instrument.  

\subsubsection*{Landform and Hydrology Metrics}
A selection of 20 terrain morphological and hydrological metrics were computed from a 25m resolution Digital Elevation Model (DEM) for the catchment in which the study area was located (see Table \ref{t:covar.acrons}) with the System for Automated Geoscientific Analyses (SAGA v2.1.0) \cite{Conrad2015}.
This selection was based on DEM products that have been used in other soil carbon modelling studies \cite{ViscarraRossel2014,Miller2016}.

\subsubsection*{Remotely Sensed Data}
Imagery from an airborne $\gamma$ ray radiometric survey has also been used to model soil organic carbon \cite{Rawlins2009}.
We were able to obtain such imagery with a 50m$^2$ resolution for our study area \cite{Brown2003} and included the Potassium, Uranium and Thorium channels along with the Total Dose among our potential covariates.
The location of woody vegetation has also been found correlated with SOC concentration \cite{Graham2004,Hibbard2011}.
Foliar Projective Cover (FPC) rasters obtained from applying the Statewide Landcover and Trees Study (SLATS) (see \url{https://www.qld.gov.au/environment/land/vegetation/mapping/slats-methodology/}) method to 2011 and 2012 imagery from the SPOT5 satellite (10m$^2$ resolution) were thus also included as potential covariates.
These layers were obtained from the New South Wales State Government Department of Environment.

\subsection*{Statistical Methods}

\subsubsection*{Interpolation of Covariates to the Spatial Neighbourhoods of Response Observations}
All covariates were obtained as either high resolution geostatistical data across the study sites from the ATV transects or as full cover rasters of various resolutions that encompassed the study sites.
We interpolated all covariates to 25m by 25m squares centred on each response observation.
Raster covariates were interpolated to these squares by taking an area based weighted average of the values of the pixels that intersected these squares.
Thin plate spline surfaces were fitted to all the ATV derived geostatistical covariates with the R package `fields' \cite{Nychka2016}.
These spline surfaces were then used to predict the values of these covariates at regularly spaced rectangular arrays of points within these 25m by 25m squares centred on the location of each response observation.
For each covariate, the average of these predicted values was taken as the value of that covariate accompanying the response observation at the center of the 25m by 25m square.
A 25m by 25m square area around each response observation was selected as the majority of the covariate rasters were calculated from the 25m$^2$ resolution DEM.

\subsubsection*{Design Matrix Construction and Treatment of Site Information}
Prior to any modelling, a subset of covariates was selected among which no pair had a correlation coefficient greater in magnitude than 0.95.
Our method of choosing members of correlated pairs of covariates for inclusion in this subset has been outlined in S2 Appendix 2.
Essentially, finer resolution covariates were chosen over coarser resolution covariates and simpler covariates were chosen over more complex covariates.
We considered polynomial terms for all covariates up to order four and all possible pairwise interactions of linear terms for covariates.
This resulted in our set of 65 potential covariates being expanded to 2340 potential covariates prior to filtering to enforce a maximum degree of collinearity among covariates.
\newline
\newline
Site specific models were fitted to training sets constructed from observations of the site in question.
We will refer to this method of modelling as Method 1 hereafter.
Models of the data combined from both sites, which we denote global effects models, were fitted to training sets containing equal numbers of observations from both sites.
We will refer to fitting models that used covariates as global effects only (that is effects constant across both sites) to the data combined from both sites as Method 2 hereafter.
We also amend the model averaged predictions from Method 2 with the predictions from site specific models fitted to the residuals from the model averaged predictions from Method 2.
We refer to this method of amending the predictions from the global effects models with the predictions from site specific models for the residual from these global effect models as Method 3.
Design matrices containing global and site specific effects were created by combining three copies of the design matrix for the global effects of covariates across the data.
The first copy was left unmodified to serve as the global effects terms for all covariates.
The site specific effects were then created by replacing entries in the second and third copies of this design matrix with zeros in the rows which corresponded to observations from the other site.
This structure is laid out in Table \ref{t:GESE.DM} where $x_{i,j,k}$ is the $i^{th}$ observation of the $j^{th}$ covariate considered either as a pooled effect across sites ($k=0$) or as a site specific effect ($k=1$ for site B1 and $k=2$ for site B2).
We refer to this method of fitting models with design matrices that contain covariates as both global effects and site specific effects as Method 4. 

\subsubsection*{Recentring and Rescaling of Covariates}
The LAR algorithm requires all input covariates to have means of zero and L2 norms of one \cite{Efron2004}.
In order to maintain the function of the site specific effects we conducted the recentring and rescaling in a manner that preserved the entries of zero in the appropriate rows of the columns for site specific effects of covariates in the design matrices.
It was necessary to recenter and rescale the design matrices for each training set individually and store these transformations so that identical transformation could be applied to the design matrices constructed from the associated validation set and to the design matrices constructed from the covariates interpolated to the pixels of the full cover raster.
The design matrices for the validation sets were constructed so that models could be assessed on the accuracy with which they predicted data held out from the fitting process as detailed in the next section on \nameref{Sec:LASSO.MLR}.
The design matrices for the full cover covariate rasters were constructed so models could provide full cover raster predictions for soil carbon as detailed in the section on \nameref{Sec:Areal.Inf}.
Further details regarding the recentring and rescaling of covariates prior to use with the LAR algorithm are provided in S1 Appendix 1.

\subsubsection*{LASSO Regularized MLR}
\label{Sec:LASSO.MLR}
The familiar MLR model predicts the observations of the response $y_i$, $i = 1, ..., n$, with an intercept term $\hat \beta_0$ plus a linear combination of products of observations of the covariates $x_{i,j}$, $j = 1, ..., p$, and the associated coefficient estimates $\hat \beta_j$ as per Equation \ref{Eq:MLR}.
\begin{equation}
\label{Eq:MLR}
\hat y_{i} = \hat \beta_0 + \sum_{j=1}^p \hat \beta_j x_{i,j}.
\end{equation}
The maximum likelihood estimate of the coefficient vector $\mathbf{ \beta } = ( \beta_0, \beta_1, ... , \beta_p )^T$ for an MLR model is obtained identifying the $\mathbf{\hat \beta}$ that minimizes the residual sum of squares as per Equation \ref{eq:MLR.beta.hat}.
\begin{equation} \label{eq:MLR.beta.hat}
\bm{\hat \beta} = \argmin\limits_{\bm{\beta}} \{ \sum\limits_{i = 1}^n (y_{i} - \beta_0 - \sum\limits_{j = 1}^px_{ij} \beta_j)^2 \}.
\end{equation}
The LASSO regularized estimate of the coefficient vector $\mathbf{ \beta }$ for an MLR model is obtained by identifying the $\mathbf{\hat \beta}$ that minimizes the sum of the residual sum of squares and a multiple of the sum of the absolute values of the coefficients (i.e. a multiple of the L1 norm of the coefficient vector $\mathbf{\beta}$) as per Equation \ref{eq:LASSO.beta.hat}.
\begin{equation} \label{eq:LASSO.beta.hat}
\bm{\hat \beta} = \argmin\limits_{\bm{\beta}} \{ \sum\limits_{i = 1}^n (y_{i} - \beta_0 - \sum\limits_{j = 1}^px_{ij} \beta_j)^2 + \lambda \sum\limits_{j = 1}^p | \beta_j | \}.
\end{equation}
The tuning parameter $\lambda$ is sometimes referred to as the shrinkage parameter and controls the degree to which $\bm{\hat \beta}$ is shrunk towards the zero vector.
It is the nature of LASSO regularization that elements of $\mathbf{\beta}$ will be shrunk to zero exactly.
The number of elements shrunk to zero depends on the value of $\lambda$.
We were interested in fitting models to predict \%SOC at each pixel of the DEM (i.e. between soil core locations) either at the same site as the data to which the model is being fitted were collected or at the other site.
Subsequently, we elected to estimate $\lambda$ via a cross validation scheme so as to select a subset size from the model choice trajectory that was best for out of sample prediction.
We used the same 500 unique divisions of the data from each site into training sets of 35 observations and validation sets of 25 (or 21) observations in each modelling scenario conducted.
Where we were fitting models to data from both sites combined, we constructed training sets by combining the training sets from the two sites and validation sets by combining the associated validation sets from both sites.
We used the LAR algorithm \cite{Efron2004} as implemented in the R package `lars' \cite{Hastie2013} to calculate LASSO solution paths.

\subsubsection*{Model Averaging}
Applying the LAR variable selection algorithm within a cross validation scheme yielded one selected model for each training set.
The predictions from each of the selected models were model averaged using a weighted average.
To emphasize out of sample predictive accuracy the weights for the model averaging, $W_i$, were inversely proportional to the validation set prediction error sum of squares as per Equation \ref{eq:MA.Weights}.
Here, $i$ indexes the 500 divisions of the data into training and validation sets and $e_{i,\,j}$ represents the prediction error of the $j^\text{th}$ element of the $i^\text{th}$ validation set of $v$ elements.
\begin{equation} \label{eq:MA.Weights}
W_i = \frac{(\sum\limits_{j=1}^{v}{e_{i,\,j}^2})^{-1}}{\sum\limits_{i=1}^{500}{({\sum\limits_{j=1}^{v}{e_{i,\,j}^2}})^{-1}}}.
\end{equation}

\subsubsection*{Model Comparison}
We compared the models selected for the considered scenarios in terms of the frequencies with which the various covariates were selected, metrics of the accuracy of the model averaged predictions including the coefficient of determination and the residual mean square error, and histograms of the residuals arising from the predictions from these models.
The frequencies with which covariates were selected across the 500 selected models (one for each training set) have been depicted via chord diagrams.
In these diagrams the covariates have been arrayed around the perimeter of a circle.
Interaction terms are represented by curved lines (Poincar\'{e} segments) joining the constituent terms of the interaction term.
Polynomial terms are represented by points of different shapes at radii passing through the appropriate covariate acronym at distances from the centre of the circle proportional to the polynomial order.
The opacity of the point or Poincar\'{e} segment is proportional to the frequency with which the respective covariate occurred in the 500 models selected for that scenario.

\subsection*{Areal Inference}
\label{Sec:Areal.Inf}
After interpolating all covariates to the pixels of the DEM, we were able to calculate model averaged predictions for \%SOC at each of these pixels from each of the sets of selected models.
To predict from each model, we used covariate rasters that had been recentred and rescaled with identical transformations to those applied to the covariates in the respective training set, then calculated prediction rasters from each of the 500 selected models.
The model averaged predictions for each pixel were then calculated from the 500 predicted values for that pixel as a weighted averaged using the weights calculated with Equation \ref{eq:MA.Weights}.

\subsection*{Software and Code}
All programming and statistical analyses were conducted in the R language and environment for statistical computing \cite{R2015}.
All graphics were produced with the R package `ggplot2' \cite{Wickham2009}.
The DEM derived covariates were calculated with the System for Automated Geoscientific Analyses (SAGA v2.1.0) \cite{Conrad2015} software and read into R with the R package `RSAGA' \cite{RSAGA2008}.
The R package `reshape' \cite{Wickham2007} was used for data manipulation and the R package `xtable' \cite{xtable2015} was used for conversion of dataframes to LaTeX tables.
The R code written for this analysis has been provided via a Git repository as detailed in S1 Code.

\section*{Results}
\label{s:res}
\subsection*{Summary of Analysis}
A panel of chord diagrams representing the frequencies with which covariates were selected from the site specific models (Method 1) is presented as Fig \ref{Fig:SS_Chord_Diag_Pannel}.
The 20 most frequently selected covariates from the B1 and B2 site specific models are presented in Tables \ref{Tab:B1.SE.T20.Covars} and \ref{Tab:B2.SE.T20.Covars} respectively.
Alongside each selected covariate in these tables are listed any other covariates that were very strongly correlated and hence were excluded from the analysis in the filtering stage.
Fig \ref{Fig:GE_Chord_Diag} and Table \ref{Tab:GE.T20.Covars} provide analogous information for the models of the data from sites B1 and B2 combined that were built from a set of consisting covariates of only global effects terms (Method 2).
Fig \ref{Fig:GESE_Chord_Diag} and Table \ref{Tab:GESE.T20.Covars} provide analogous information for the models of the data from sites B1 and B2 combined that were built from a set of covariates that included both global effects terms and site specific effects terms (Method 3).
We refer to the number of covariates in a selected model as the subset size.
The distributions of selected subset sizes across the 500 selected models from each of Methods 1 to 4 have been presented as boxplots with the individual values overlaid as horizontally jittered points in Fig. \ref{Fig:Sel_Sub_Set_Size}.
For Methods 1 to 4, the coefficients of determination ($R^2$) for the model averaged predictions from each of these methods are presented in Table \ref{t:R2}.
The residual mean squared error (RMSE) values associated with the model averaged predictions from Methods 1 to 4 have been presented in Table \ref{t:RMSE}. 
Histograms of the residuals from these model averaged predictions have been presented in Fig. \ref{Fig:Resid_Histograms_Tech_x_Site}.

\subsection*{Method 1: Site Specific Models}
The $R^2$ value for the model averaged predictions of the B2 response from models selected for the B2 data was almost identical to that for the model averaged predictions of the B1 response from models selected for the B1 data (see Table \ref{t:R2}).
However, the Residual Mean Squared Error (RMSE) associated with these predictions (see Table \ref{t:RMSE}) revealed that the model averaged predictions of the B2 data from the models fitted to these data to be much more accurate than the model averaged predictions of the B1 data from the models fitted to the B1 data.
Model averaged predictions from the models selected for the data from site B1 with Method 1 proved more geographically transferable than did model averaged predictions from the models selected for the data from site B2 with Method 1.
This may be seen in the contrasting accuracies with which the response observations at each site were predicted from the observations of the covariates at that site using the models fitted to the data from the other site.
The model averaged predictions of the B1 response from the appropriately recentred and rescaled B1 covariates using the models fitted to the data from site B2 had a negative $R^2$ and an RMSE of 0.89.
In contrast, the model averaged predictions of the B2 response from the appropriately recentred and rescaled covariate observations from site B2 using the models fitted to the data from site B1 had an $R^2$ of 0.30 and an RMSE of 0.23.

\subsection*{Method 2: Global Effects Models}
The model averaged predictions from the global effects models (Method 2) predicted the response observations more accurately than either of the site specific models (Method 1), see the $R^2$ values in Table \ref{t:R2} and the corresponding RMSE values in Table \ref{t:RMSE}. 
This difference was substantial for the data from site B1 and much smaller but still noticeable in data from site B2.
Furthermore, prediction with the site specific models (Method 1) resulted in more extreme residuals than prediction with the global effects models (see Fig \ref{Fig:Resid_Histograms_Tech_x_Site}).
Of the ten most frequently selected covariates among the global effects models, only the Difference Vegetation Index calculated from the May top of pasture reflectance values (`May.DVI') was present in both the most frequently selected covariates terms from the global effects models and the most frequently selected covariates from the site specific models.
The models specific to site B1 included far more intercept only models than the global effects models fitted to the data combined from both sites (see Fig. \ref{Fig:Sel_Sub_Set_Size}).
This however was not the case among the models specific to site B2 (see Fig. \ref{Fig:Sel_Sub_Set_Size}).
This was notable as the the global effect models had much better predictive accuracy compared to the site specific models for site B1 while a similar improvement was not observed over the models specific to the data from site B2.

\subsection*{Methods 3 \& 4: Global and Site Specific Effects Models}
The two staged approach to incorporating site specific effects into models of the data combined from both sites (Method 3) resulted in more accurate prediction of the observed data than the approach whereby global effects and site specific effects were considered together in a single round of modelling (Method 4), see Tables \ref{t:R2} and \ref{t:RMSE}.
The predictions from Method 4 also resulted in more extreme residuals than the predictions from the two staged approach, see Fig \ref{Fig:Resid_Histograms_Tech_x_Site}.
The predictions from the global effects only models (Method 2) were also more accurate than the predictions from Method 4, see Tables \ref{t:R2} and \ref{t:RMSE} and Fig \ref{Fig:Resid_Histograms_Tech_x_Site}.

\subsection*{Areal Inference}
With all covariates interpolated to a common raster we were able to calculate model averaged predictions for each pixel in this raster with each of the Methods 1 to 4.
The rasters predicted by Method 1 (the site specific models) have been presented in Fig. \ref{Fig:MAP_Rast_SS}.
The raster predicted by Method 3 (the global effects only model amended with the site specific models) has been presented in Fig. \ref{Fig:MAP_Rast_GEpSS}.
The raster predicted by Method 2 (the unamended, global effects only model) has been presented in S4 Fig 1 and the raster predicted from Method 4 (global effects and site specific effects) has been presented in S5 Fig 1.

\section*{Discussion}
\label{s:disc}
We have explored the geographic transferability of site specific models and compared different implementations of site specific effects into models for data combined from two sites via a digital soil mapping case study.
The computational efficiency of the LAR algorithm for LASSO regularized MLR enabled linear, polynomial and pairwise linear interaction terms to be considered for all covariates in each of the models we have fitted.
Furthermore, we were able to fit and compare models with different treatments of site specific effects for all covariates (linear, polynomial and interaction terms) some of which required design matrices with three times the number of columns used in the design matrices for simpler approaches.

\subsection*{Site Specific Models and Geographic Transferability}
\begin{figure}[h!]
\label{Fig:Covar_Dist_Comp_Box_Plots}
\includegraphics[width = \textwidth]{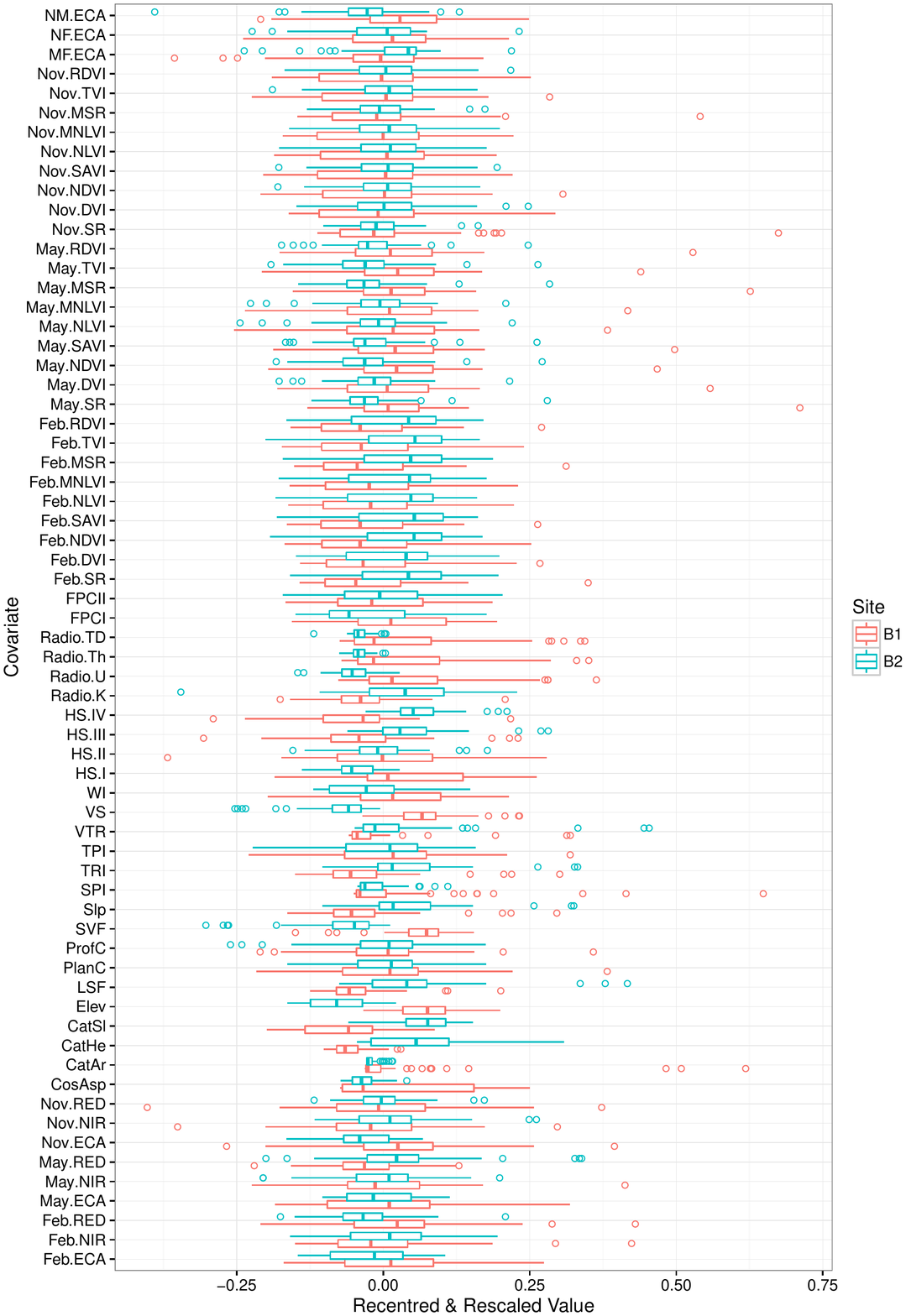}
\caption{Boxplots comparing the distributions of the covariates at either site. The observations compared are the covariate values interpolated to the locations at which soil cores were collected. The sets of observations of each covariate pooled across both sites have each been recentred to have means of zero and rescaled to have magnitudes of one.} 
\end{figure}

\begin{figure}[h!]
\label{Fig:Hist_Reponse}
\includegraphics[height = 0.5\textheight]{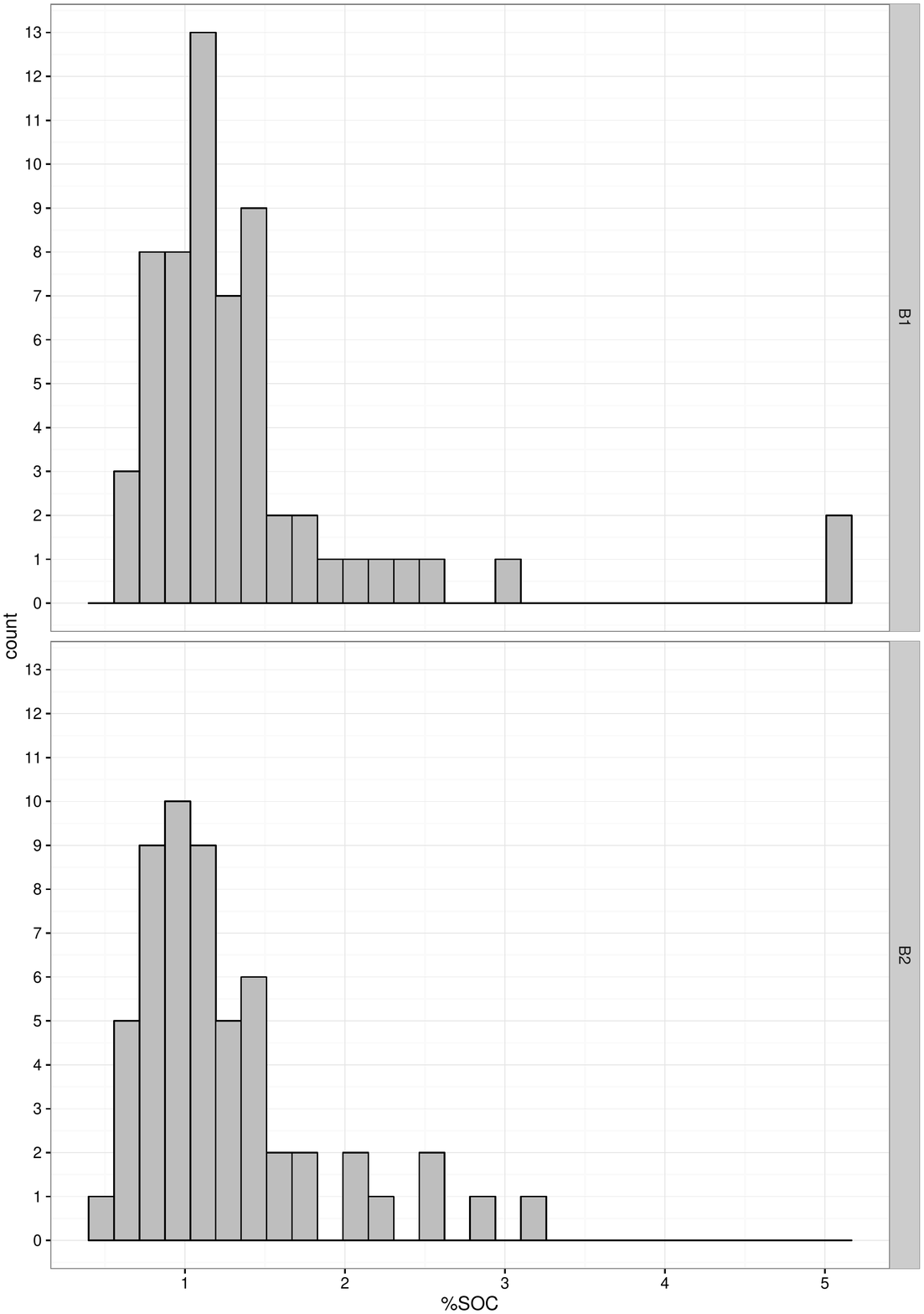}
\caption{A panel of histograms of the percentage soil organic carbon (\%SOC) observations obtained from each site. The upper histogram depicts the distribution of soil carbon observations from site B1 and the lower histogram depicts the distribution of the soil carbon observations from site B2.}
\end{figure}

\begin{figure}[h!]
\label{Fig:SS_Chord_Diag_Pannel}
\includegraphics[width = \textwidth]{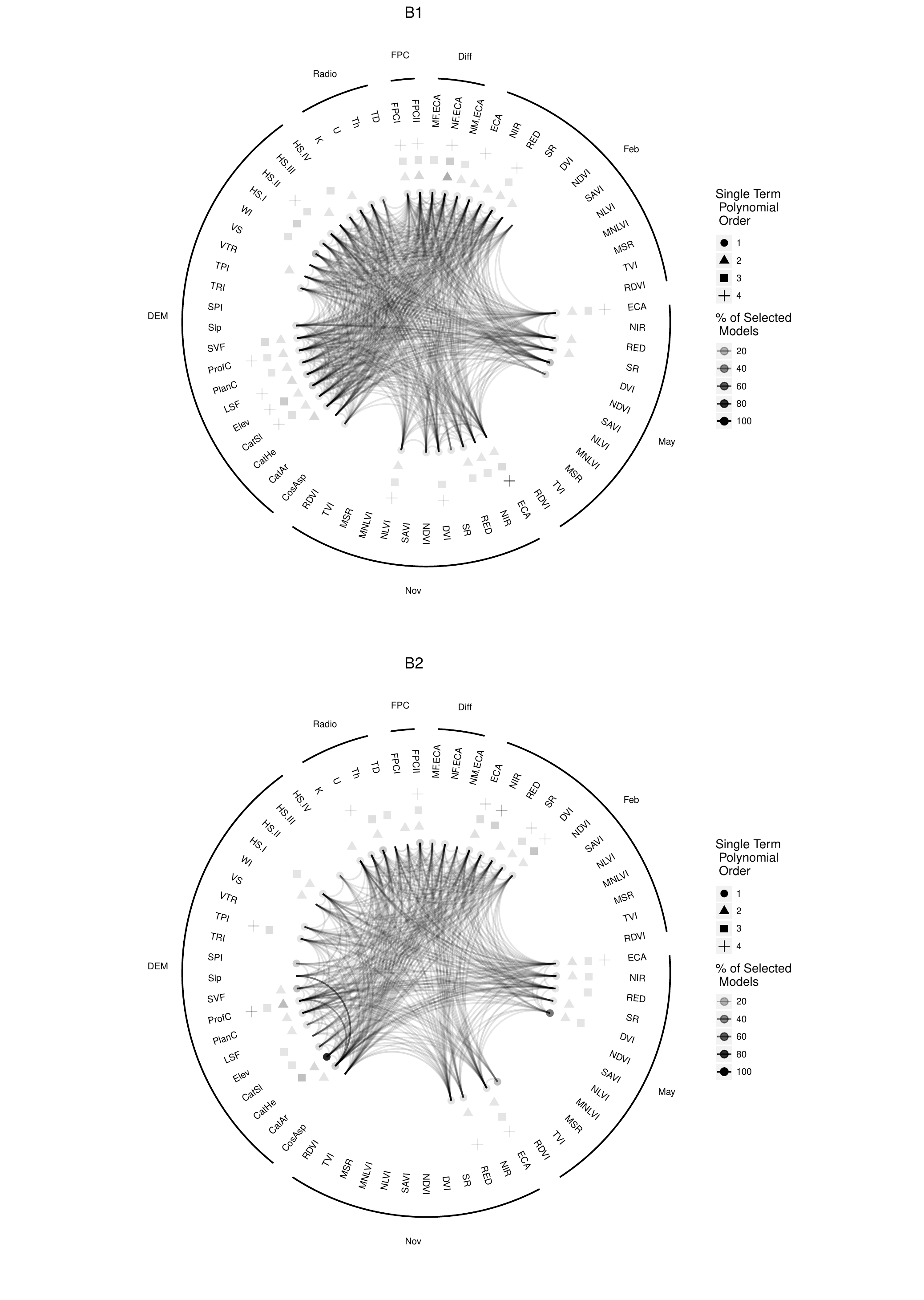}
\caption{\textbf{a)} A chord diagram depicting the frequencies of selection of the covariates from the site specific modelling of the data from site B1 \textbf{b)} A chord diagram depicting the frequencies of selection of the covariates terms from the site specific modelling of the data from site B2.}
\end{figure}

\begin{figure}[h!]
\label{Fig:GE_Chord_Diag}
\includegraphics[width = \textwidth]{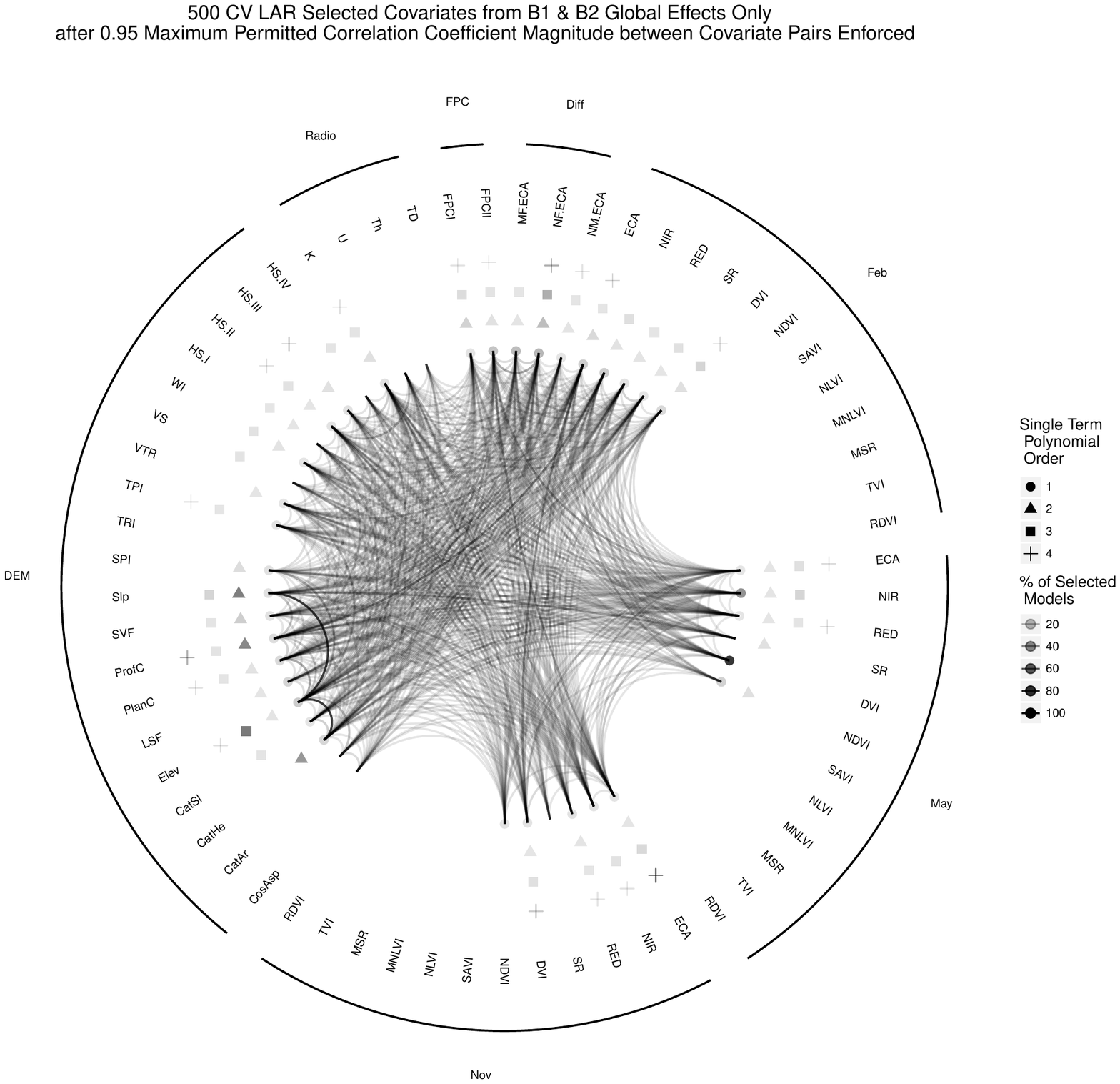}
\caption{A chord diagram depicting the frequencies of selection of the covariates from the modelling of the pooled data from sites B1 and B2 that considered covariates as global effects only.}
\end{figure}

\begin{figure}[h!]
\label{Fig:GESE_Chord_Diag}
\includegraphics[width = \textwidth]{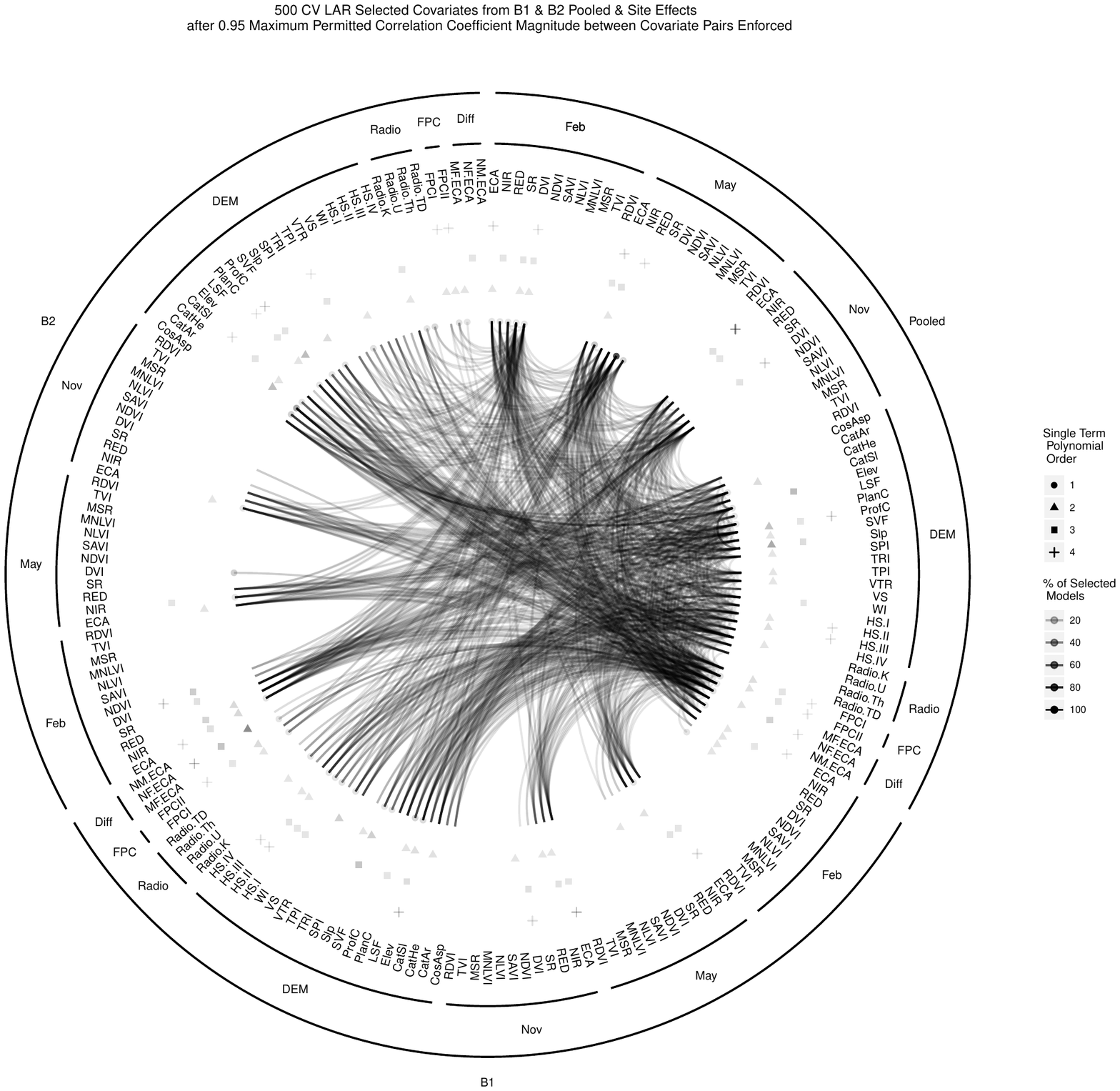}
\caption{A chord diagram depicting the frequencies of selection of the covariates in the modelling of the pooled data from sites B1 and B2 that considered covariates as both Global Effects and Site Specific Effects.}
\end{figure}

\begin{figure}[h!]
\label{Fig:Sel_Sub_Set_Size}
\includegraphics[width = 0.35\textwidth]{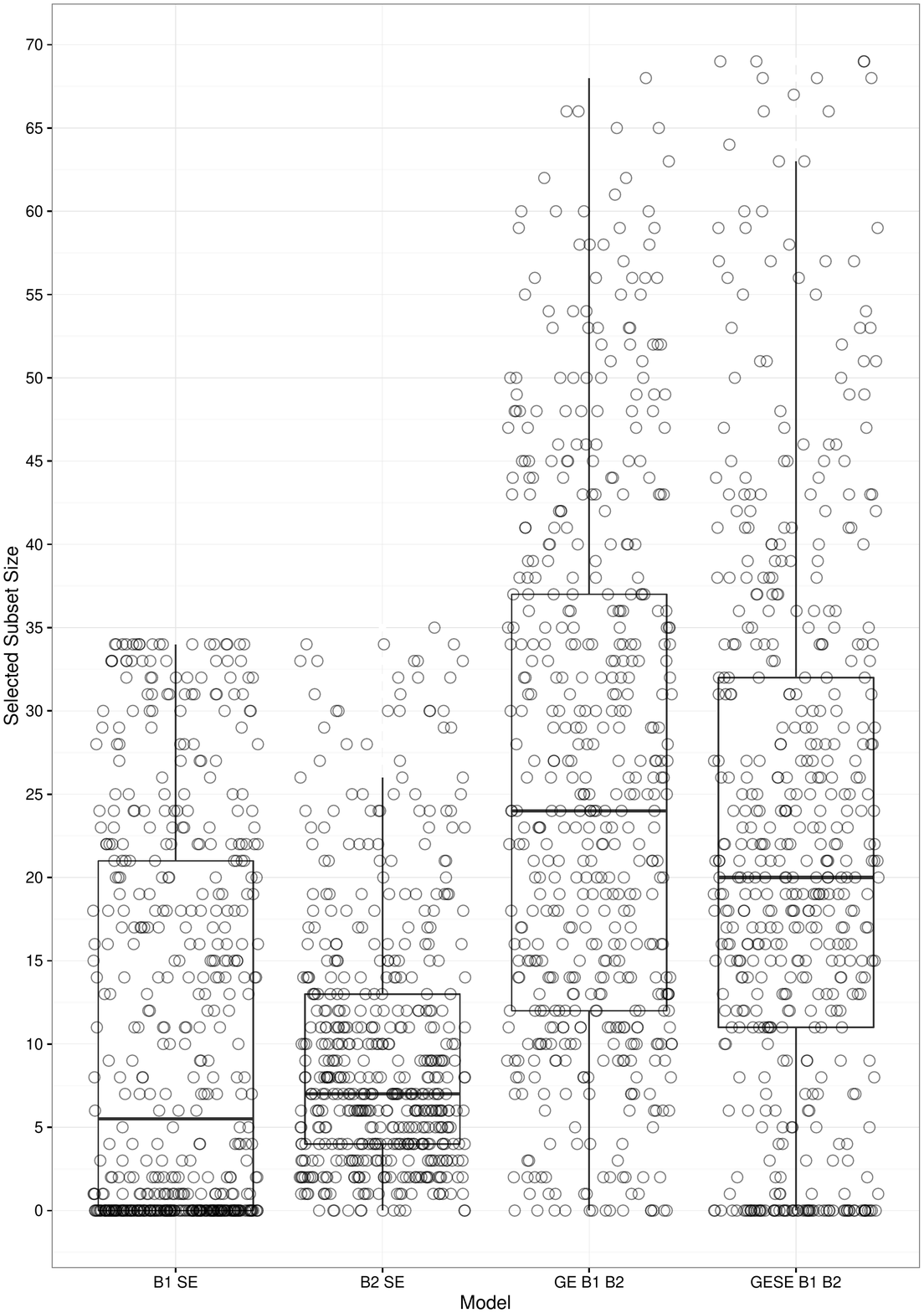}
\caption{Boxplots of the selected subset sizes from the different modelling methods considered with the individual observations overlaid as horizontally jitter points. Selected subset sizes of zero equate to intercept only models. B1 SE = Site Specific Models for data from Site B1, B2 SE = Site Specific Models for data from Site B2, GE B1 B2 = Global Effects (only) models for data pooled from sites B1 and B2, GESE B1 B2 = Global Effect and Site Specific Effects models for data pooled from sites B1 and B2.}
\end{figure}

\begin{figure}[h!]
\label{Fig:Resid_Histograms_Tech_x_Site}
\includegraphics[width = 0.35\textwidth]{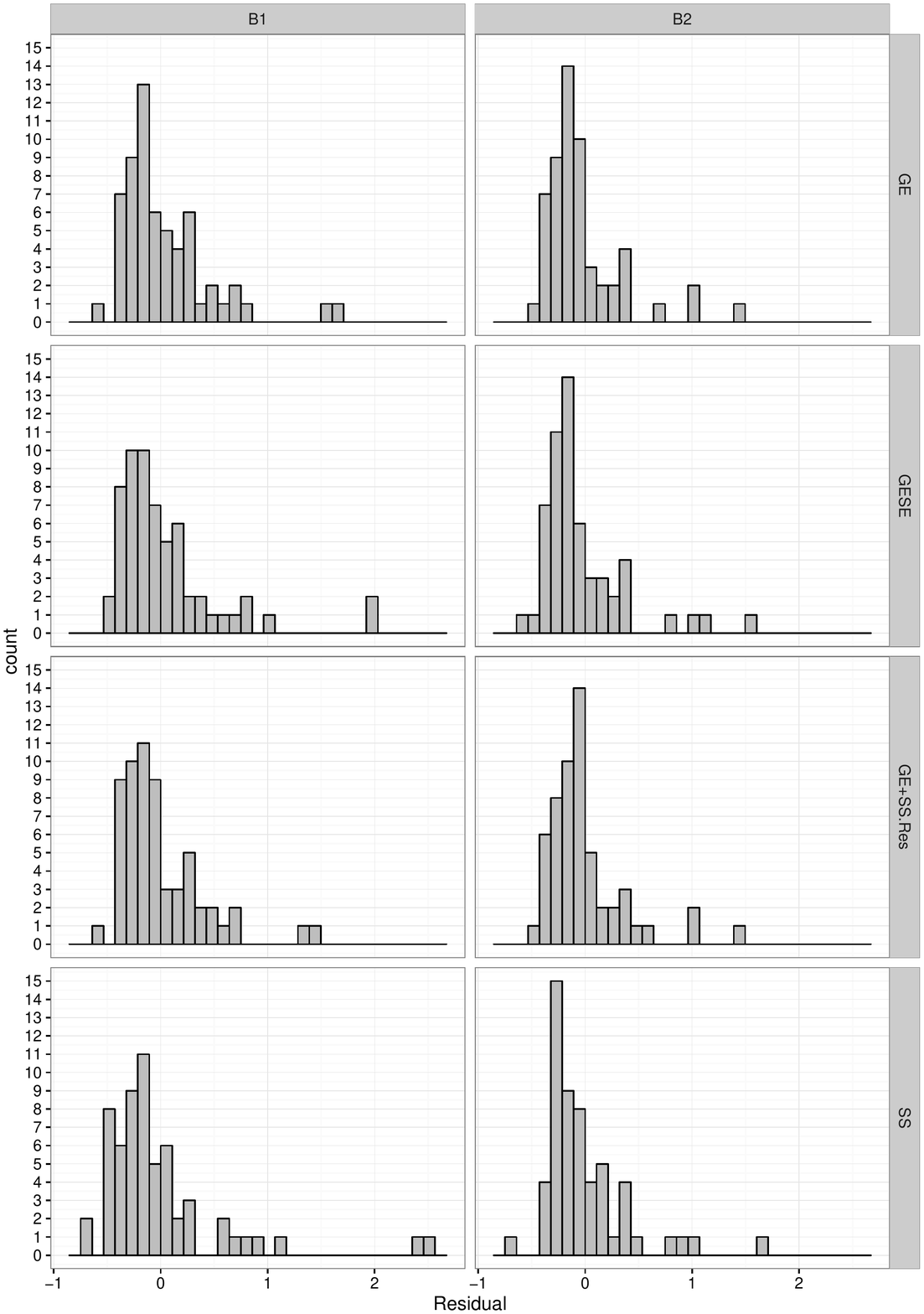}
\caption{Histograms of the residuals from the model averaged predictions from each of the modelling approaches trialed at each site. GE = Global Effect (only) models. GESE = Global Effects and Site Specific Effects models. GE+SS.Res = Global Effects models with prediction ammended by predictions from Site Specific models of the residuals from the global effects model. SS = Site Specific Models.}
\end{figure}

\begin{figure}[h!]
\label{Fig:MAP_Rast_SS}
\includegraphics[width = \textwidth]{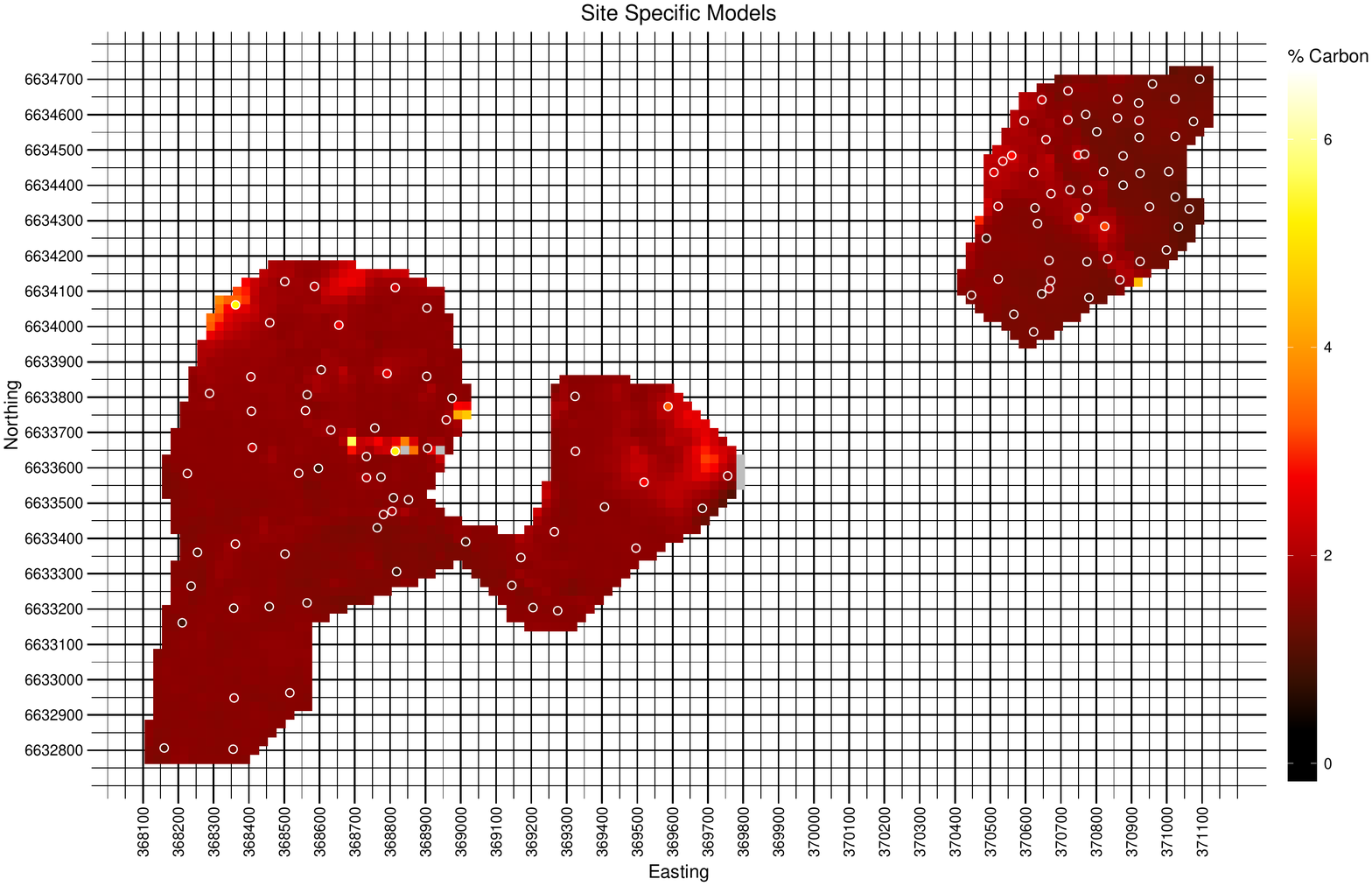}
\caption{The model averaged prediction rasters for both sites from the site specific models. Raster pixels are coloured proportionally to the the model averaged prediction for that pixel. The locations at which soil cores were collected have been depicted as black circles and the colour filling each soil core is proportional to the soil carbon value obtained from that soil core. Grey pixels have positive values greated than the maximum value in the colour scale used.}
\end{figure}

\begin{figure}[h!]
\label{Fig:MAP_Rast_GEpSS}
\includegraphics[width = \textwidth]{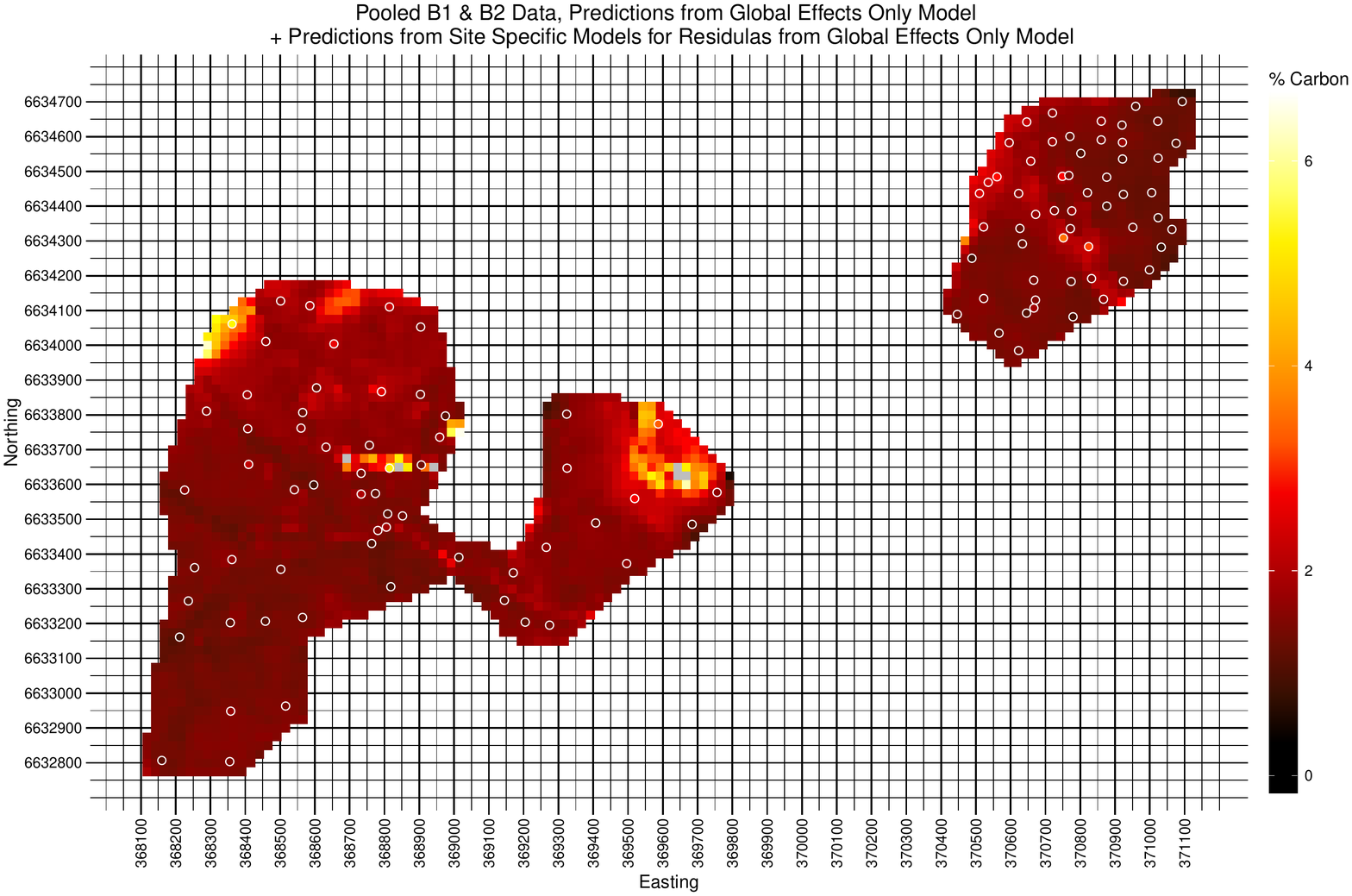}
\caption{The model averaged prediction rasters for both sites from the global effects model amended with the predictions from the site specific models. Raster pixels are coloured proportionally to the the model averaged prediction for that pixel. The locations at which soil cores were collected have been depicted as black circles and the colour filling each soil core is proportional to the soil carbon value obtained from that soil core. Grey pixels have positive values greated than the maximum value in the colour scale used.}
\end{figure}

The contrast between the $R^2$ and RMSE values for the model averaged predictions of the data at each site with the site specific models (Method 1) fitted to those data is likely the result of the two positive outliers in the response at Site B1 (see Fig. \ref{Fig:Hist_Reponse}) since the $R^2$ statistic is strongly affected by outliers.
The accuracies of the spatial extrapolations of the site specific models from one site to the other site were used to investigate the geographic transferability of these site specific models.
Unsurprisingly, spatial extrapolation of site specific models from one site to the other site was more accurate when it involved less extrapolation with respect to the covariates most important for predicting the response in these site specific models.
\newline
\newline
Catchment Height, the interaction between Catchment Area and land surface Slope and the Difference Vegetation Index from May were the three most frequently selected covariates among the models selected for the B2 data (see Table \ref{Tab:B2.SE.T20.Covars}).
Prediction of site B1 response observations from site B1 covariate observations with models fitted to site B2 data would have been gross extrapolation in terms of each of these three most frequently selected covariates among the models selected for the data from site B2 (see \ref{Fig:Covar_Dist_Comp_Box_Plots}).
By far the most frequently selected covariate among the 500 selected models for the data from site B2 was the Catchment Height (`CatHe') which was selected in over 80\% of these models.
However, almost 50\% of the Catchment Height values interpolated to the soil core locations at site B1 were below the range of Catchment Height values used in fitting the models to the data from site B2.
This meant that the spatial extrapolation of the B2 models to the B1 data was also a major extrapolation with respect to this covariate.
The next most frequently selected covariate among the models selected for the B2 data was the interaction between Catchment Area and land surface Slope (`CatAr:Slp') which was selected in just over half of these models.
Fig \ref{Fig:Covar_Dist_Comp_Box_Plots} depicts how the top 25\% of the Catchment Area values interpolated to the soil core locations at site B1 were greater than the largest value among the interpolations of this covariate to locations of the soil cores at site B2.
Furthermore, the interquartile range of these CatAr values at site B2 was less than half that at site B1.
Similarly, the central 50\% of the observations of land surface Slope at site B1 were all less than the central 50\% of the observations of land surface slope at site B2.
The implication again is that gross extrapolation was being carried out with respect to this covariate when the models fitted to the data from site B2 were used to predict the B1 response from the B1 covariates.
The Difference Vegetation Index calculated from the pasture reflectance data collected in May (May.DVI) was also selected in just over half of the models selected for the B2 data. 
The middle 50\% of the May DVI values at site B2 occupied approximately half of the width of the May DVI values from site B1 so again prediction using the covariate May DVI at site B1 using models fitted to the data from site B2 would have been extrapolation with respect to this covariate in many cases.
\newline
\newline
The three most frequently selected covariates among the models selected for the data from site B1 were: 1) the power four polynomial for the soil apparent electrical conductivity from the November survey (`Nov.ECA'), 2)  the quadratic term for the difference between the soil apparent electrical conductivity measurements from November and February (`NF.ECA.2') and 3) the soil Wetness Index (`WI') (see Table \ref{Tab:B1.SE.T20.Covars}).
The ranges of the interpolated values of Nov.ECA and WI at the soil core locations at site B1 encompassed the respective ranges of the interpolated values of these covariates at the soil core locations at site B2.
The range of NF.ECA values at site B2 was contained (with the exception of a single positive outlier) within the range of the observations of this covariate at site B1.
Thus the geographic extrapolation conducted when assessing the transferability to site B2 of the models fitted to the data from site B1 was not extrapolation with respect to the three most frequently selected covariates in these models.
Whereas, the geographic extrapolation conducted when models fitted to data from site B2 were used to predict the response at site B1 from the covariates at site B1 was a case of severe extrapolation in terms of the most frequently selected covariates among these models.
Thus in our digital soil mapping case study analysis, spatial extrapolation was more successful when it involved less extrapolation in terms of the covariates most important for predicting the response (see Tables \ref{t:RMSE} and \ref{t:R2}). 
This provides a reminder of the hazards of spatial extrapolation to even spatially proximate locations.

\subsection*{Site Specific Models Compared to Models for the Data from both Sites Combined}
The models fitted to data from both sites combined predicted the B1 response observations more accurately than the site specific models fitted to the B1 data alone.
The models fitted to the data from both sites combined predicted the B2 response observations with similar or greater accuracy to the site specific models fitted to the B2 data alone.
However, this difference was much more substantial in the case of site B1 than site B2.
The models that were specific to the site B1 data (Method 1 applied to Site B1) had many more intercept only models among them than the global effects models (Method 2) that were fitted to the data from sites B1 and B2 combined.
Whereas, the models specific to site B2 (Method 1 applied to Site B2) had a similar number of intercept only models among them compared to the global effects models (Method 2) that were fitted to the data from both sites combined. 
This may have influenced the improved prediction accuracy of the global effects models over the site B1 specific models on the site B1 data.
The two positive outliers in the response at site B1 may have skewed many of the fits of the site specific models there such that the best model for predicting the associated validation set from those model selection trajectories was an intercept only model.
Such models, however, would not have generalized well.
In contrast, these two positive outliers at site B1 would have had far less of an effect on the model fitting in the global effects models of the combined data since these models benefited from training sets twice the size of the training sets used in fitting the site specific models to data from either site alone.
Support for this interpretation may be found in how of the ten most frequently selected covariates among the global effects models, only May.DVI was present among the ten most frequently selected covariates from each of the site specific models.

\subsection*{Two Stage Treatment of Site Specific Effects Compared to Single Treatment of Global and Site Specific Effects}
We have referred to the approach of fitting global effects only models first then explaining the residual variation with site specific models (Method 3) as a two staged approach.
When modelling the data combined from both sites, more accurate predictions were obtained from models that took a two staged approach to incorporating site specific effects (Method 3) than from models where global effects and site specific effects were considered together in a single round of modelling (Method 4).
One possible explanation stems from how the LAR algorithm iteratively chooses covariates to add to the current model based on which of the covariates absent from the current model is most correlated with the current residual vector.
As such LAR is by no means guaranteed to identify the `best' model for a particular set of data.
Instead, LAR iteratively builds models by selecting the `best' covariate to add to the model at each particular iteration.
While iteratively improving the current model compared to model from the previous iteration, this strategy is by no means guaranteed to return the globally optimal model when the algorithm halts.
The approach considering global effects and site specific effects resulted in more intercept only models being selected by cross validation from the model choice trajectories output by the LAR algorithm than were selected among the global effect models that did not consider site specific effects.
Thus it is possible that the LAR algorithm followed more model selection trajectories that were sub-optimal for predicting the associated validation sets under the global effects and site specific effects scenario than under the global effects only scenario.
However, it is also possible that fitting a global effects only model on data pooled across sites where covariates acted in the same manner at both sites more realistically reflected the underlying processes influencing the observed soil carbon distribution across these two sites.
Fitting global effects only models first restricted the LAR algorithm to building models from such covariate effects that were common across sites thereby avoiding the potential for the algorithm to add to the model site specific effects that at particular steps better explained the current residual vector but ultimately resulted in a model with poorer predictive performance.
The site specific covariate effects could then be used in the second stage of the modelling to explain residual variation in the soil carbon from this first modelling stage.
If the computational challenges were able to be met perhaps Bayesian LASSO \cite{Park2008} or Bayesian Spike and Slab variable selection \cite{Ishwaran2005} on the global effects and site specific effects scenario would perform better than LAR variable selection as these techniques would converge towards to high posterior probability models rather than proceeding to deterministically build models based on covariate correlations with residual vectors as the LAR algorithm does.

\subsection*{Conclusion}
Our digital soil mapping case study highlights the importance of the manner in which site specific effects are incorporated into models of data combined from multiple sites.
In this case study, our site specific models had greater geographic transferability when the spatial extrapolation involved was accompanied by less extrapolation in the important covariates in these models.
While specific to our data, our results still highlight the importance of considering these issues when modelling data from multiple sites and or seeking to extrapolate spatially.
As data from multiple sites are shared and used in an aggregated fashion these concerns become pertinent to modelling in an increasing number of fields.

\section*{Acknowledgements}
The work has been supported by the Cooperative Research Centre for Spatial Information (CRCSI), whose activities are funded by the Australian Commonwealth's Cooperative Research Centres Programme.
One of the authors (BRF) wishes to acknowledge the receipt of a Postgraduate Scholarship from the CRCSI.
We thank Vincent Zoonekynd for the R code to calculate Poincar\'{e} segment paths.

\section*{Tables}
\begin{table}[h!]
\caption{The covariates used in the analysis.}
\label{t:covar.acrons}
\begin{tabular}{lll}
Category    & Acronym  & Full Name \\
\hline
\hline
In-situ     & ECA       & Soil Apparent Electrical Conductivity ($EC_a$) \\
Surveys:    & NIR       & Near InfraRed Reflectance \\
Feb, May,   & RED       & Red Reflectance \\
\& Nov      & SR        & Simple Ratio \\
            & DVI       & Difference Vegetation Index \\
            & NDVI      & Normalized Difference Vegetation Index \\
            & SAVI      & Soil Adjusted Vegetation Index \\
            & NLVI      & Non-Linear Vegetation Index \\
            & MNLVI     & Modified Non-Linear Vegetation Index \\
            & MSR       & Modified Simple Ratio \\
            & TVI       & Transformed Vegetation Index \\
            & RDVI      & Re-normalized Difference Vegetation Index \\
\hline
In-situ     & MF.ECA    &  May $EC_a$ - Feb $EC_a$\\
Survey      & NF.ECA    &  Nov $EC_a$ - Feb $EC_a$\\
Differences & NM.ECA    &  Nov $EC_a$ - May $EC_a$\\
\hline
Digital     & CatAr     & Catchment Area \\
Elevation   & CatHe     & Catchment Height \\
Model       & CatSl     & Catchment Slope \\
Derived     & CosAsp    & Cosine(Aspect) \\
            & Elev      & Elevation \\
            & LSF       & Slope Length Factor \\
            & PlanC     & Plan Curvature \\
            & ProfC     & Profile Curvature \\
            & SVF       & Sky View Factor \\
            & Slp       & Slope \\
            & SPI       & Stream Power Index \\
            & TRI       & Terrain Ruggedness Index \\
            & TPI       & Topographic Position Index \\
            & VTR       & Vector Terrain Ruggedness \\
            & VS        & Visible Sky \\
            & WI        & Wetness Index \\
            & HS.I      & Hill Shading (Angle I) \\
            & HS.II     & Hill Shading (Angle II) \\
            & HS.III    & Hill Shading (Angle III) \\
            & HS.IV     & Hill Shading (Angle IV) \\
\hline
Remotely    & Radio.K   &  $\gamma$ Radiometric Potassium \\
Sensed      & Radio.U   &  $\gamma$ Radiometric Uranium \\
            & Radio.Th  &  $\gamma$ Radiometric Thorium \\
            & Radio.TD  &  $\gamma$ Radiometric Total Dose \\
            & FPCI      &  Foliar Projective Cover 2011 \\
            & FPCII     &  Foliar Projective Cover 2012 \\
\end{tabular}
\end{table}

\begin{table}[h!] 
\caption{Structure of the design matrix for the Global Effect and Site Specific Effects models. $x_{i,j,k}$ is the $i^{th}$ observation of the $j^{th}$ covariate considered either as a pooled effect across sites ($k=0$) or as a site specific effect (for site B1 $k=1$, or for site B2 $k=2$). Obs. = Observation Number}
\label{t:GESE.DM}
\begin{tabular}{l c c c c c c c c c c}
Site & Obs.& \multicolumn{3}{c}{Global Effects} & \multicolumn{3}{c}{Site B1 Effects} & \multicolumn{3}{c}{Site B2 Effects} \\
B1   & 1   & $x_{1,1,0}$   & ... & $x_{1,2340,0}$   & $x_{1,2341,1}$ & ... & $x_{1,4680,1}$   & 0              & ... & 0            \\
     &     & .           & ... &   .            &              & ... & .              &                & ... &              \\
     &     & .           & ... &   .            &              & ... & .              &                & ... & .            \\
     &     & .           & ... &   .            &              & ... & .              &                & ... & .            \\
B1   & 60  & $x_{60,1,0}$  & ... & $x_{60,2340,0}$  & $x_{60,2341,1}$ & ... & $x_{60,4680,1}$  & 0              & ... & 0             \\
B2   & 1   & $x_{61,1,0}$  & ... & $x_{61,2340,0}$  & 0            & ... & 0              & $x_{61,4681,2}$   & ... & $x_{61,7020,2}$ \\
     &     & .           & ... & .              & .            & ... & .              & .              & ... & .             \\
     &     & .           & ... & .              & .            & ... & .              & .              & ... & .              \\
     &     & .           & ... & .              & .            & ... & .              & .              & ... & .              \\
B2   & 56  & $x_{116,1,0}$ & ... & $x_{116,2340,0}$  & 0            & ... & 0              & $x_{116,4681,2}$ & ... & $x_{116,7020,2}$  \\
\end{tabular}
\end{table}

\begin{table}[h!] 
\caption{The coefficients of determination associated with the model averaged predictions from the different models fitted.}
\label{t:R2}
\begin{tabular}{ c c c c c c}
         &           &           & Fitted to        & Fitted to        & Fitted to \\
         &           &           & B1 \& B2         & B1 \& B2         & B1 \& B2 \\
         &           &           & Data             & Data             & Data (Global \\
         & Fitted to & Fitted to & (Global          & (Global \&       & Effects + Site\\
Predict  & B1 Data   & B2 Data   & Effects)         & Site Effects)    & Specific Models)  \\
\hline
\hline
B1       &   0.51    &  -0.29*   &      0.74        &  0.66            & 0.78 \\
B2       &   0.30    &   0.52    &      0.58        &  0.51            & 0.61 \\
B1 \& B2 &    NA     &    NA     &      0.70        &  0.61            & 0.73 \\
\hline
\multicolumn{6}{l}{*i.e. considerably worse than an intercept only model} \\
\end{tabular}
\end{table}

\begin{table}[h!] 
\caption{The residual mean square errors values associated with the model averaged predictions from the different models fitted.}
\label{t:RMSE}
\begin{tabular}{ c c c c c c}
         &           &           & Fitted to        & Fitted to        & Fitted to \\
         &           &           & B1 \& B2         & B1 \& B2         & B1 \& B2 \\
         &           &           & Data             & Data             & Data (Global \\
         & Fitted to & Fitted to & (Global          & (Global \&       & Effects + Site\\
Predict  & B1 Data   & B2 Data   & Effects)         & Site Effects)    & Specific Models)  \\
\hline
\hline
B1       &  0.34     &  0.89     &  0.18          & 0.24                    & 0.15 \\
                 &           &           &                &                         &      \\
B2       &  0.23     &  0.16     &  0.14          & 0.16                    & 0.12 \\
                 &           &           &                &                         &      \\
B1 \& B2 &    NA     &    NA     &  0.16          & 0.20                    & 0.14 \\
\hline
\multicolumn{6}{l}{*i.e. considerably worse than an intercept only model} \\
\end{tabular}
\end{table}

\begin{table}[h!] 
\caption{A summary of the covariate selection frequencies from the site specific models fitted to the data from site B1. In particular, the twenty most frequently selected covariates, the frequency of selection of these covariates and the correlated covariates that were filtered out in order to retain each of these covariates.} 
\label{Tab:B1.SE.T20.Covars}
\begin{tabular}{lrl}
  \hline
Covariate         & Freq & Correlated Covariates \\ 
  \hline
  Nov.ECA.4       &  179 & \\ 
  NF.ECA.2        &  122 & \\ 
  WI              &   95 & \\ 
  Slp:May.SR      &   89 & Slp:May.NDVI, Slp:May.MSR, Slp:May.TVI, \\
                  &      & TRI:May.SR, TRI:May.NDVI, TRI:May.MSR \\ 
  May.DVI         &   87 & May.SAVI, May.NLVI, May.MNLVI, May.RDVI \\ 
  HS.II:Nov.DVI   &   77 & HS.II:Nov.MNLVI  \\ 
  ProfC:HS.I      &   74 & \\ 
  FPCII:NF.ECA    &   72 & \\ 
  Elev            &   64 & \\ 
  May.RED:CatHe   &   58 & \\ 
  Nov.ECA:Nov.DVI &   57 & Nov.ECA:Nov.MNLVI \\ 
  Elev.3          &   56 & \\ 
  CatAr:Radio.U   &   54 & SPI:Radio.U \\ 
  NF.ECA.3        &   53 & \\ 
  May.RED:WI      &   52 & \\ 
  May.ECA:FPCII   &   49 & \\ 
  Elev:Slp        &   47 & Elev:TRI\\ 
  Feb.ECA:SVF     &   47 & \\ 
  CatSl.4         &   46 & \\ 
   \hline
\end{tabular}
\end{table}

\begin{table}[h!]
\caption{A summary of the covariate selection frequencies from the site specific models fitted to the data from site B2.  In particular, the twenty most frequently selected covariates, the frequencies of selection of these covariates and the correlated covariates that were filtered out in order to retain these covariates.}
\label{Tab:B2.SE.T20.Covars}
\begin{tabular}{lrl}
  \hline
Covariate & Freq & Correlated Covariates \\ 
  \hline
  CatHe &   416 &  \\ 
  CatAr:Slp &   246 & CatAr:TRI, CatAr:HS.IV \\ 
  May.DVI &   246 & May.NLVI, May.MNLVI \\ 
  CatAr:ProfC &   149 &   \\ 
  Feb.NIR.4 &   110 &   \\ 
  May.RED:CatAr &   100 &  \\ 
  ProfC.2 &    90 &   \\ 
  ProfC.4 &    86 &   \\ 
  May.RDVI &    84 & May.SAVI  \\ 
  SVF &    83 & VS  \\ 
  CatHe.3 &    81 & CatHe.4   \\ 
  May.ECA:Radio.K &    73 &   \\ 
  May.RED:Radio.TD &    70 &  \\ 
  Feb.DVI.3 &    68 & Feb.MNLVI.3  \\ 
  FPCII &    63 &   \\ 
  May.RED:ProfC &    62 &    \\ 
  Radio.Th &    59 &    \\ 
  CatAr:Radio.TD &    54 &    \\ 
  CatAr &    53 &    \\ 
  May.RED:Radio.K &    46 &   \\ 
   \hline
\end{tabular}
\end{table}

\begin{table}[h!]
\caption{A summary of the covariates selected for the global effects only models fitted to the data from sites B1 and B2 combined. In particular, the 20 most frequently selected covariates, the frequencies with which these covariates were selected and the correlated covariates that were filtered out in order to retain these covariates.} 
\label{Tab:GE.T20.Covars}
\begin{tabular}{lrl}
  \hline
Covariate         & Freq & \\ 
  \hline
  May.DVI         &   381 & May.NLVI, May.MNLVI, May.RDVI \\ 
  CatHe:Elev      &   366 & \\ 
  Elev:Slp        &   366 & Elev:TRI \\ 
  Nov.ECA.4       &   273 & \\ 
  Elev.3          &   234 & \\ 
  Slp.2           &   224 & Slp:TRI, Slp:VTR, TRI:VTR, TRI.2 \\ 
  ProfC.2         &   201 & \\ 
  May.NIR         &   183 & \\ 
  Feb.NIR:ProfC   &   169 & ProfC:Feb.DVI, ProfC:Feb.MNLVI \\ 
  CatHe.2         &   168 & CatHe.3, CatHe.4 \\ 
  Feb.ECA:Nov.DVI &   142 & Feb.ECA:Nov.MNLVI, Feb.ECA:Nov.RDVI \\ 
  Elev:PlanC      &   127 & \\ 
  Slp:May.SR      &   124 & Slp:May.NDVI, Slp:May.SAVI, Slp:May.MSR, \\
                  &       & Slp:May.TVI, TRI:May.SR, TRI:May.NDVI,  \\ 
                  &       & TRI:May.SAVI, TRI:May.MSR, TRI:May.TVI \\
  FPCII:NF.ECA    &   121 &\\ 
  Nov.ECA:FPCII   &   121 &\\ 
  NF.ECA.3        &   118 & \\ 
  Elev:SPI        &   109 & \\ 
  NF.ECA.4        &   107 & \\ 
  ProfC.4         &   105 &   \\ 
  NF.ECA          &   101 & \\ 
   \hline
\end{tabular}
\end{table}

\begin{table}[h!]
\caption{A summary of the covariate selection frequencies of the global effects and site specific effects models fitted to the data from sites B1 and B2 combined.  In particular, the twenty most frequently selected covariates, the frequencies with which these covariates were selected and the correlated covariates that were filtered out in order to retain these covariates.} 
\label{Tab:GESE.T20.Covars}
\begin{tabular}{lrl}
  \hline
Covariate & Freq & Correlated Covariates \\ 
  \hline
May.DVI &   341 & May.NLVI, May.MNLVI, May.RDVI  \\ 
  Elev:Slp &   231 & Elev:TRI\\ 
  Nov.ECA.4 &   230 & B1.Nov.ECA.4 \\ 
  B2.CatAr:Elev &   195 & \\ 
  B1.NF.ECA.2 &   194 &\\ 
  CatHe:Elev &   191 & \\ 
  B2.CatHe &   131 &\\ 
  Slp.2 &   125 & Slp:TRI, Slp:VTR, TRI:VTR, TRI.2  \\ 
  B1.FPCII:NF.ECA &   121  \\ 
  B1.Elev:WI &   113 & \\ 
  Nov.ECA:FPCII &   110 & \\ 
  B2.CatHe:MF.ECA &   109 &\\ 
  B1.HS.II:Nov.DVI &    90 & B1.HS.II:Nov.MNLVI \\ 
  B1.NF.ECA.4 &    89 & \\ 
  Elev.3 &    88 &  \\ 
  B2.ProfC.2 &    87 &\\ 
  B1.Nov.ECA.4 &    85 & Nov.ECA.4\\ 
  B2.CatHe.2 &    81 & CatHe.2, CatHe.3, CatHe.4, B2.CatHe.3, B2.CatHe.4 \\ 
  B1.May.RED:WI &    79 &  \\ 
  B2.Elev:HS.IV &    79 &   \\ 
   \hline
\end{tabular}
\end{table}
\clearpage

\section*{Figures}

\clearpage

\section*{Supporting Information}

\begin{figure}[h!]
\label{S1_Fig_Aerial_Photo_B1}
\includegraphics[width = \textwidth]{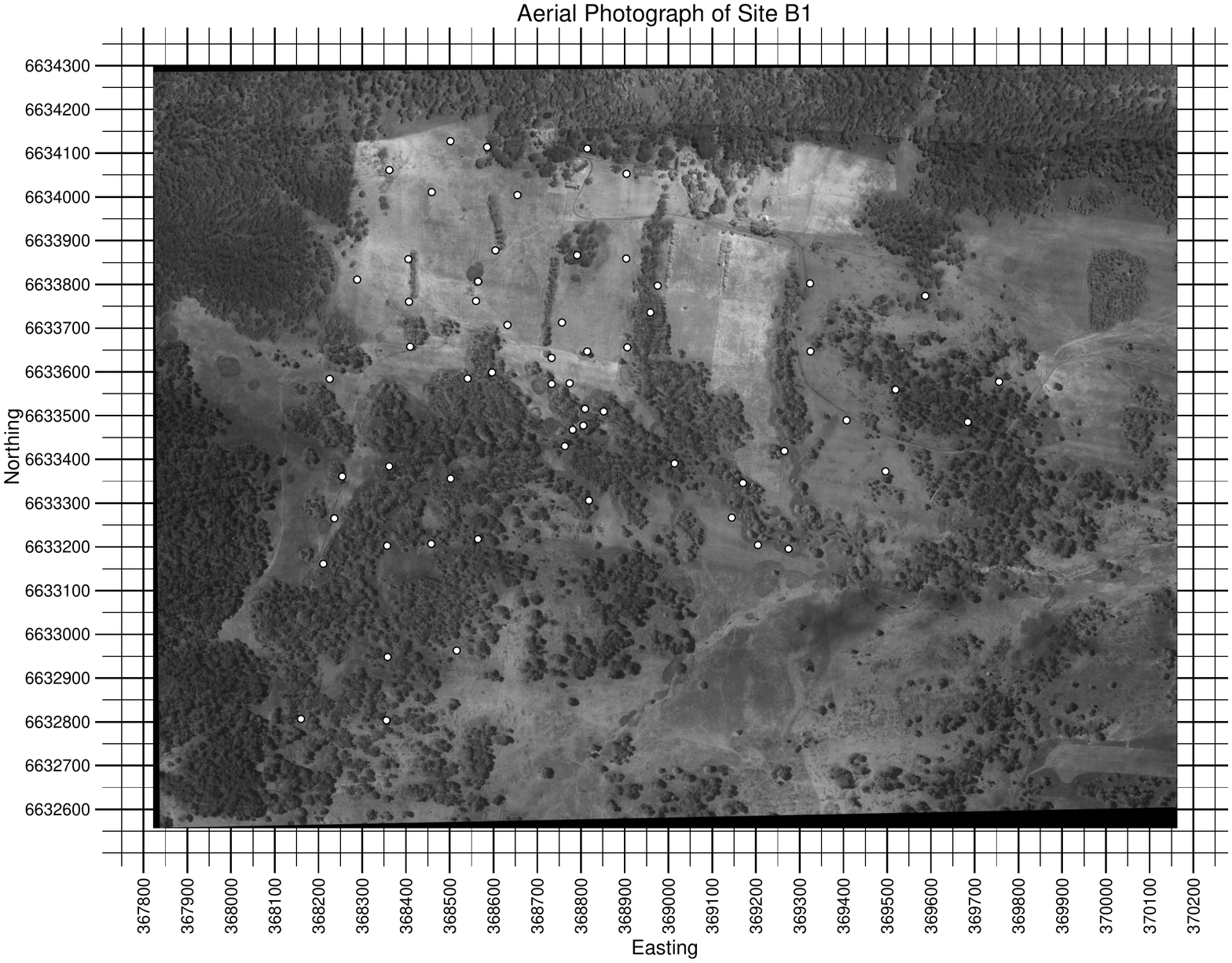}
\caption{S1 Figure: Aerial photograph of site B1 with soil core locations marked as white filled black circles.}
\end{figure}

\begin{figure}[h!]
\label{S2_Fig_Aerial_Photo_B2}
\includegraphics[width = \textwidth]{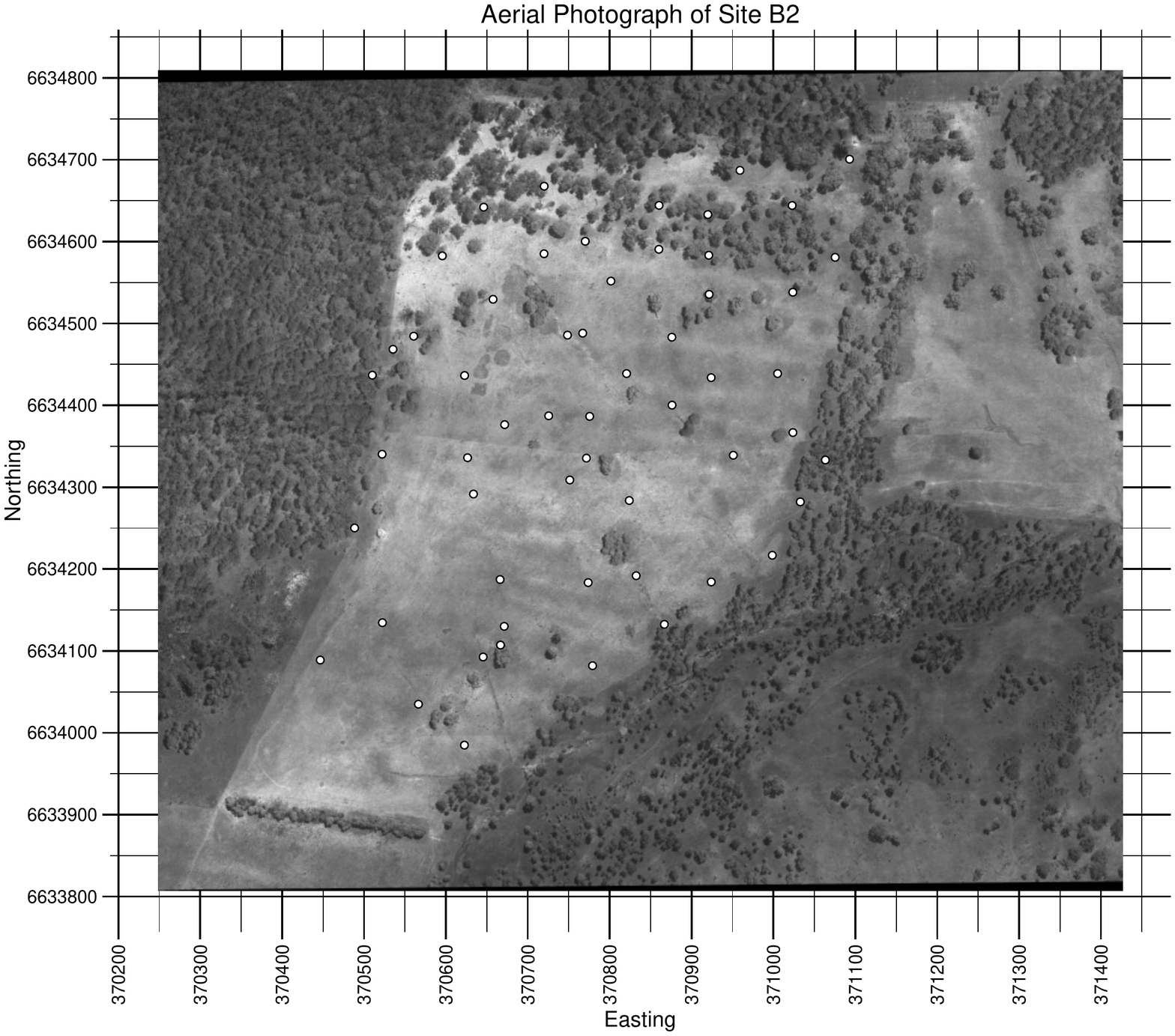}
\caption{S2 Figure: Aerial photograph of site B2 with soil core locations marked as white filled black circles.}
\end{figure}

\begin{figure}[h!]
\label{S3_Fig_MAP_Rast_GE}
\includegraphics[width = \textwidth]{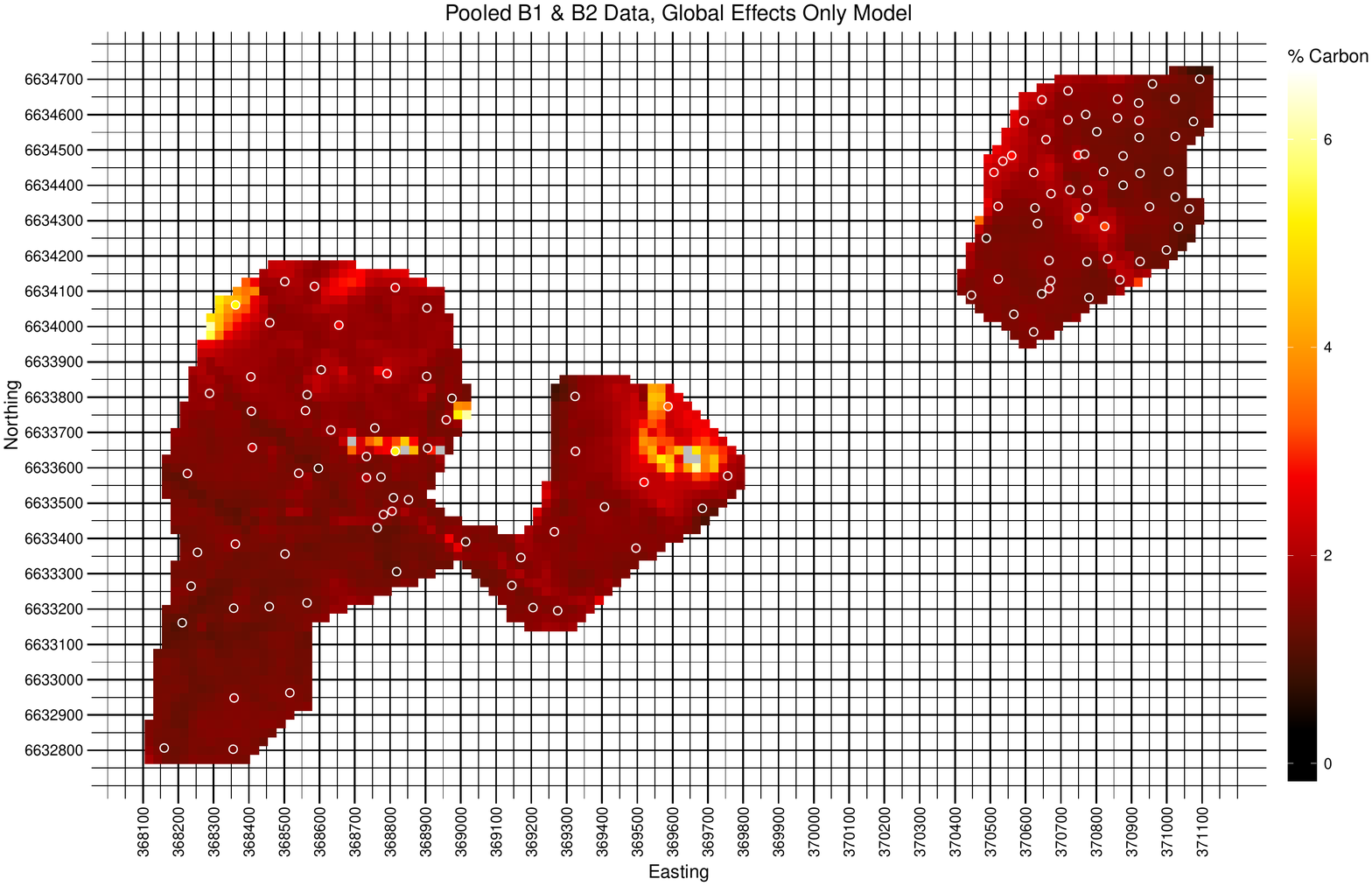}
\caption{S3 Figure: The model averaged prediction rasters for both sites from the global effects only model of combined data from both sites. Raster pixels are coloured proportionally to the the model averaged prediction for that pixel. The locations at which soil cores were collected have been depicted as black circles and the colour filling each soil core is proportional to the soil carbon value obtained from that soil core. Grey pixels have positive values greated than the maximum value in the colour scale used.}
\end{figure}

\begin{figure}[h!]
\label{S4_Fig_MAP_Rast_GESE}
\includegraphics[width = \textwidth]{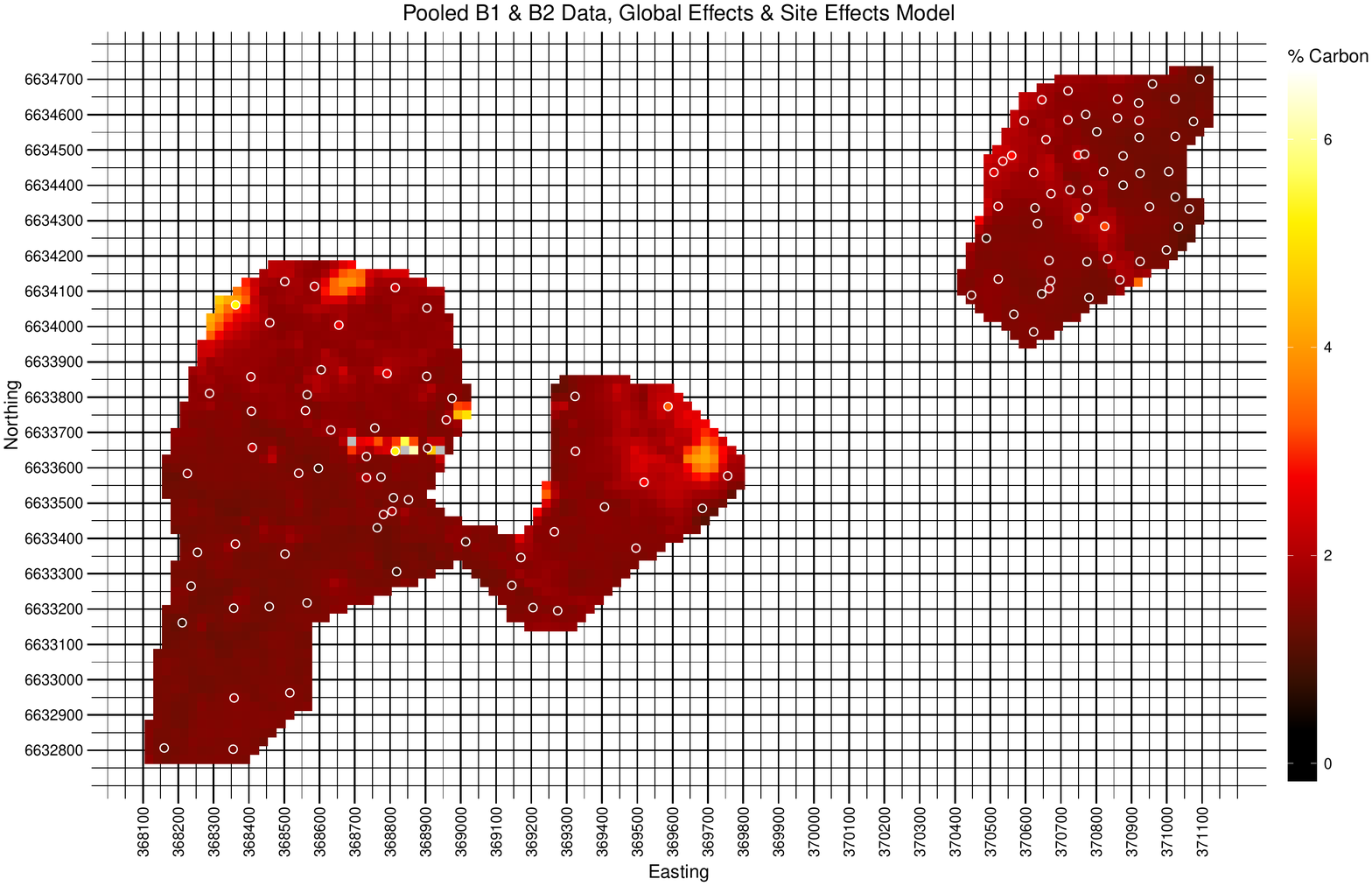}
\caption{S4 Figure: The model averaged prediction rasters for both sites from the global effects and site specific effects model of combined data from both sites. Raster pixels are coloured proportionally to the the model averaged prediction for that pixel. The locations at which soil cores were collected have been depicted as black circles and the colour filling each soil core is proportional to the soil carbon value obtained from that soil core. Grey pixels have positive values greated than the maximum value in the colour scale used.}
\end{figure}
\clearpage

\subsection*{S1 Appendix 1}
\label{S1_Text}
\textbf{Appendix detailing the recentring and rescaling of covariates necessary for each modelling method.}
\newline
\newline
The Least Angle Regression (LAR) algorithm \cite{Efron2004} for calculating LASSO penalized estimates of the coefficients in multiple linear regression models relies on two assumptions regarding the covariates.
To explain these assumptions let the $i$th observation of the $j$th covariate be denoted as $x_{ij}$ for observations $i \in \{1, ..., n\}$ and covariates $j \in \{1,...,p\}$.
The first assumption of the LAR algorithm is that each covariate ($x_{.j}$) has been recentred to have a mean of zero (Equation \ref{Eq:Covar.Mean.Zero}) ($x_{ij}^{\prime}$ being an observation of the recentred covariate $x_{.j}$).
\begin{equation}
\label{Eq:Covar.Mean.Zero}
\sum_{i=1}^n x_{ij}^{\prime} = 0
\end{equation}
The second assumption of the LAR algorithm is that each covariate has been rescaled to have a magnitude of one (Equation \ref{Eq:Covar.Mag.One}) ($x_{ij}^{\prime \prime}$ being an observation of the rescaled covariate $x_{.j}$).
\begin{equation}
\label{Eq:Covar.Mag.One}
 \sum_{i=1}^n ( x_{ij}^{\prime \prime} )^2 = 1
\end{equation}
\newline
\newline
When cross validation is used to estimate the shrinkage parameter $\lambda$ in a LASSO fit (see Equation 3 in the main text) care must be taken to ensure that the covariate observations that comprise each validation set are recentred and rescaled with transformations identical to those used to recentre and rescale the covariate observations in each associated training set.
For simplicity of notation, assume that of our $n$ observations, the first $t$ observations constitute a training set , $\{1, ... , t\}$, and the remaining $n-t$ observations, $\{t+1, ... , n\}$, constitute the associated validation set.
Prior to using the LAR algorithm to estimate a LASSO solution the observations of the training set need to be recentred to have a mean of zero with Equation \ref{Eq:Train.Covar.Mean.Zero} and then rescaled to have a magnitude of one with Equation \ref{Eq:Train.Covar.Mag.One}.
\begin{equation}
\label{Eq:Train.Covar.Mean.Zero}
x_{ij}^\prime = x_{ij} - \frac{1}{t}\sum_{i=1}^t x_{ij}, i \in \{1,...,t\}, \forall j
\end{equation}

\begin{equation}
\label{Eq:Train.Covar.Mag.One}
x_{ij}^{\prime \prime} = \frac{x_{ij}^{\prime}}{(\sum_{i=1}^t (x_{ij}^{\prime})^2)^{\frac{1}{2}}}, i \in \{1,...,t\}, \forall j
\end{equation}
To use the model selected for a training set to make predictions with the data from the associated validation set, the covariate observations from the validation set must be recentred and rescaled with the same transformations that were applied to the covariate observation in the training set (as per Equations \ref{Eq:Valid.Covar.Mean.Train} and \ref{Eq:Valid.Covar.Mag.Train}).

\begin{equation}
\label{Eq:Valid.Covar.Mean.Train}
x_{ij}^\prime = x_{ij} - \frac{1}{t}\sum_{i=1}^t x_{ij}, i \in \{t+1,...,n\}, \forall j
\end{equation}

\begin{equation}
\label{Eq:Valid.Covar.Mag.Train}
x_{ij}^{\prime \prime} = \frac{x_{ij}^{\prime}}{(\sum_{i=1}^t (x_{ij}^{\prime})^2)^{\frac{1}{2}}}, i \in \{t+1,...,n\}, \forall j
\end{equation}

To use the model fitted to a training set to make predictions with covariate observations from a different site to that from which the training set observations were collected, the observations from the other site need to be recentred and rescaled with the same transformations that were applied to the covariate observation from the training set.
That is the observation from the other site need to be recentred and rescaled with $ \frac{1}{t}\sum_{i=1}^t x_{ij}$ and $(\sum_{i=1}^t (x_{ij}^{\prime})^2)^{\frac{1}{2}}$ where $i \in \{1,...,t\}$ index the observations in the training set.
\newline
\newline
It is necessary to recenter and rescale the covariate observations being used to make predictions with transformations identical to those applied to the covariate observations to which the model in question was fitted so that the predictions are constructed from covariate observations on the same (artificial) scales as those from which the coefficient estimates were calculated.
These scales will be unique to each covariate in each training set.
Furthermore, $(\sum_{i=1}^t (x_{ij}^{\prime})^2)^{\frac{1}{2}}$ is a sum of squares rather than a mean sum of squares and as such will vary with training set size.
Thus such transformation of a vector of covariate observations, $x_{.j}$, from the validation set (or other site) to have mean zero and magnitude one would result in these covariate observations being on potentially quite different scales to those from which the coefficient estimates were constructed if the number of observations of in the validation set (or at the other site) was different to the number of observations in the training set.
Think here of training sets of 35 observations and rasters of approximately 2200 cells at which we wish to predict the response, where each $x_{.j}$ is observed on the same scale  (i.e. with the same units) in the training set and at the cells of the raster.
Rescaling the $x_{ij}$ in the training set by dividing each observation by $(\sum_{i=1}^{35} (x_{ij}^{\prime})^2)^{\frac{1}{2}}$ ($i$ representing observations in the training set) and rescaling the $x_{ij}$ in the raster by dividing each observation by $(\sum_{i=1}^{2200} (x_{ij}^{\prime})^2)^{\frac{1}{2}}$ ($i$ indexing the cells of the raster)  would result in very different covariate values being multiplied by the $\hat{\beta}_j$ to create the predicted response raster than were used to estimate the $\hat{\beta}_j$ from the training set.

\subsection*{S2 Appendix 2}
\label{S2_Text}
\textbf{Appendix describing the design matrix filtering heuristics.}
\newline
\newline
Substantial collinearity existed among the potential covariates.
We created subsets of the full design matrix by discarding members of highly correlated pairs of covariates in order to explore less collinear sets of covariates.
We did this to ease the computational burden involved in the variable selection in a manner that reduced the amount of information lost from the design matrix from discarding covariates.
Since we were conducting variable selection to build models for interpolation between geostatistical response observations we frequently made the choice of which covariate to retain from a highly correlated pair of covariates based on which of these covariates had observations available at a finer spatial resolution than the other.
Where both members of the correlated pair of covariates were available at the same spatial resolution we chose between them based on the potential relevance these covariates had to soil carbon distributions as informed by the soil carbon modelling literature.
Where this was not clear, we chose between correlated pairs of covariates based on which was a simpler function of the observed data.
When there was no clear choice available the selection was made a random.
The hierarchy of preference by which we chose members of highly correlated pairs of covariates to retain is summarised in Table \ref{tab:Prefilt.Heur} and explained in greater detail below.
\newline
\newline
The covariates derived from the All Terrain Vehicle (ATV) surveys had the finest spatial resolution of all the covariates considered in our case study.
In an effort to build models that would interpolate the response with the greatest spatial accuracy, when faced with highly correlated pairs of covariates we chose to retain the covariates collected by the ATV surveys over any others.
The ATV survey measured visible Red reflectance (RED), Near InfraRed reflectance (NIR) and soil apparent electrical conductivity (ECA) in each of three months.
The study site had (southern hemisphere) summer dominated rainfall and the February survey followed a week of no rain in what was otherwise the second wettest month of the year.
The May survey was conducted after a week of heavy rain while the November survey was conducted at the end of the drier winter growing season.
The RED and NIR values were used to calculate the vegetation indices whereas the ECA values were used only to calculate pairwise differences between these values from the different surveys.
Thus our first choice from any correlated pair of covariates was always an ECA value.
The pairwise differences between the ECA values from the different months were considered due to the potential these differences had to indicate changes in soil moisture which in turn could have been related to \%SOC levels via the influence SOC may exert on the infiltration of soil by water and the retention of water by soil \cite{Franzluebbers2002}.
Of the three possible pairwise differences in ECA, that between ECA from November and ECA from February was likely the most strongly related to changes in soil moisture so we elected to retain that covariate over any other covariate with which it was highly correlated.
The pairwise difference in ECA between November and May was likely the next most closely related to changes in soil moisture so we placed that difference second in our hierarchy of covariates to preferentially retain from correlated pairs of covariates.
Next in the hierarchy came the pairwise difference in ECA between the May and February survey and then any ECA value from a single month.
As the vegetation indices were theoretically more indicative of green biomass than raw RED or NIR reflectance, and thus potentially more closely related to SOC levels, vegetation indices were retained over the raw reflectance values where any such pairs were highly correlated.
\newline
\newline
After the ATV survey derived covariates the covariates with the next finest spatial resolution were those from the Foliar Projective Cover (FPC) Layers.
Thus we elected to retain these covariates over any coarser resolution covariates with which they were highly correlated.
We obtained two data such layers: the projected foliage cover for 2011 (FPCI) and the projected foliage cover for 2012 (FPCII).
Since 2011 was less temporally removed from the 2009 soil survey than 2012, FPCI was set to be preferentially retained over FPCII or any other highly correlated covariates.
\newline
\newline
Following the FPC layers in the order of spatial resolutions were the covariates calculated from the Digital Elevation Model (DEM).
Thus we preferentially retained these covariates over any coarser resolution covariates.
These data included elevation along with terrain and soil hydrology metrics calculated from the elevation.
We considered that these terrain and soil hydrology metrics came closer than raw elevation values to describing the landscape processes that may have influenced SOC formation, mineralization and or transport and thus the spatial distribution of SOC levels.
Subsequently, we elected to retain terrain and hydrology metrics over elevation should elevation have been highly correlated with any of these metrics.
We chose between highly correlated terrain and hydrology metrics at random.
\newline
\newline
The coarsest spatial resolution data we had were derived from the airborne $\gamma$ ray radiometric survey.
Thus these came last in our hierarchy of covariates to preferentially retain from highly correlated pairs of covariates.
As we have included the values related to Potassium, Uranium and Thorium along with the sum of these values, the total dose, we elected to preferentially retain Potassium, Uranium or Thorium values over the total dose should any such pairing be highly correlated.
Following the enforcement of the hierarchy of preference outlined above any remaining highly correlated pairs of covariates were chosen between at random.
\newline
\newline
Once this hierarchy of filtering operations had been applied to the 65 potential covariate terms the remainder was expanded to include polynomial terms for all remaining covariates up to polynomial order four and all possible interactions between pairs of linear terms for these remaining covariates.
In the spirit on avoiding unnecessary complexity, when searching the expanded design matrix for correlated pairs of covariate terms, single term polynomial terms (including linear terms) were retained in preference to any interaction terms with which they were found to be highly correlated.
Next, lower order polynomial terms were retained in preference to any higher order polynomial terms with which they were highly correlated.
Once all the above heuristics had been implemented in the order described a selection was made from any remaining pairs of highly correlated covariates at random to complete the enforcement of a maximum permitted correlation coefficient magnitude between covariates in the filtered design matrix.

\begin{table}[h!]
\caption{The filtering heuristics by which individual covariates were chosen from correlated pairs of covariates for retention in the filtered design matrices.}
\label{tab:Prefilt.Heur}
\begin{tabular}{ p{1cm} p{6cm} p{4cm} }
Step & Retain & Discard \\
\hline
\hline
1  & Difference between $EC_a$ in November \& February & Any Correlated Term \\
   &        &        \\
2  & Difference between $EC_a$ in November \& May & Any Correlated Term \\
   &        &        \\
3  & Difference between $EC_a$ in May and February & Any Correlated Term \\
   &        &        \\
4  & Any $EC_a$ value from a single month & Any Correlated Term \\
   &        &        \\
5  & Any Vegetation Index & Any Correlated Term \\
   &        &        \\
6  & Red reflectance or Near Infrared reflectance & Any Correlated Term \\
   &        &        \\
7  & A Foliar Projective Cover Layer & Any Correlated Term \\
   &        &        \\
8  & Any of the DEM derived covariates & Any Correlated Term \\
   &        &        \\
9  & Elevation value from the DEM & Any Correlated Term \\
   &        &        \\
10 & Radiometric Potassium, Uranium or Thorium  & Any Correlated Term \\
   &        &        \\
11 & Random & Random \\
   &        &        \\
12 & \multicolumn{2}{ l }{Expand remaining terms to include all possible polynomial terms} \\
   & \multicolumn{2}{ l }{to order 4 and all pairwise linear interactions} \\
   &        &        \\
13 & Single Term Polynomial & Interaction term for pair of covariates \\
   &        &        \\
14 & Lower Order Single Term Polynomial & Higher Order Single Term Polynomial  \\
   &        &        \\
15 & Random & Random \\
\hline
\hline
\end{tabular}
\end{table}
\clearpage
\subsection*{S1 Data}
The data analyzed in this work will be provided as a data file readable by the R language and environment for statistical computing once the article is published in a peer reviewed journal.

\subsection*{S1 Code}
The code written to conduct the analysis described in this work will be provided via a GitHub Repository once the article is published in a peer reviewed journal. The code is written in the R language for statistical computing.  The GitHub repository will be located at \url{https://github.com/brfitzpatrick/gtlarc}.

\bibliography{FLM.bib}

\end{document}